\documentclass[useAMS,usenatbib]{mn2e}

\usepackage{graphicx}
\usepackage{fixltx2e}  
\graphicspath{{astroph-figures/}}

\newcommand{\text}[1]{\mathrm{#1}}
\newcommand{\aspectratio}{\mathcal{R}}
\newcommand{\cs}{c_{\mathrm s}}
\newcommand{\jet}{{\mathrm j}}
\newcommand{\amb}{{\mathrm a}}
\newcommand{\pol}{{\mathrm p}}
\newcommand*{\avg}[1]{\langle{#1}\rangle}

\title[Very Light Magnetized Jets]{Very Light Magnetized Jets on Large Scales - I. Evolution and Magnetic Fields}
\author[V. Gaibler, M. Krause and M. Camenzind]{
V. Gaibler,$^1$\thanks{E-mail: v.gaibler@lsw.uni-heidelberg.de}
M. Krause$^{2,3}$ and M. Camenzind$^1$\\
$^{1}$Landessternwarte, Zentrum f\"ur Astronomie, Universit\"at
    Heidelberg, K\"onigstuhl~12, 69117 Heidelberg, Germany\\
$^{2}$Max-Planck-Institut f\"ur extraterrestrische Physik, Giessenbachstra\ss{}e 1, 
    85748 Garching, Germany\\
$^{3}$Universit\"ats-Sternwarte, Ludwig-Maximilians-Universit\"at (USM),
    Scheinerstr.~1, 81679 M\"unchen, Germany
}
\begin{document}

\date{Accepted 2009 August 26. Received 2009 August 12; in original form 2009 February 06}

\pagerange{\pageref{firstpage}--\pageref{lastpage}} \pubyear{2009}

\maketitle

\label{firstpage}

\begin{abstract}
  Magnetic fields, which are undoubtedly present in extragalactic jets and responsible for
  the observed synchrotron radiation, can affect the morphology and dynamics of
  the jets and their interaction with the ambient cluster medium.
  We examine the jet propagation, morphology and magnetic field structure for a
  wide range of density contrasts, using a globally consistent setup for both
  the jet interaction and the magnetic field.
  The MHD code NIRVANA is used to evolve the simulation, using the
  constrained-transport method. The density contrasts are varied between $\eta
  = 10^{-1}$ and $10^{-4}$ with constant sonic Mach number 6. The jets are
  supermagnetosonic and simulated bipolarly due to the low jet densities and their  
  strong backflows. The helical magnetic field is largely confined to the jet, leaving
  the ambient medium nonmagnetic.
  We find magnetic fields with plasma $\beta \sim 10$ already stabilize and widen the
  jet head. Furthermore they are efficiently amplified by a shearing mechanism
  in the jet head and are strong enough to damp Kelvin--Helmholtz
  instabilities of the contact discontinuity. The cocoon magnetic fields are
  found to be stronger than expected from simple flux conservation and
  capable to produce smoother lobes, as found observationally.
  The bow shocks and jet lengths evolve self-similarly. The radio cocoon aspect
  ratios are generally higher for heavier jets and grow only slowly (roughly
  self-similar) while overpressured, but much faster when they approach pressure
  balance with the ambient medium. In this regime, self-similar models can no
  longer be applied. Bow shocks are found to be of low excentricity for very
  light jets and have low Mach numbers. Cocoon turbulence and a dissolving bow
  shock create and excite waves and ripples in the ambient gas. Thermalization
  is found to be very efficient for low jet densities. 
\end{abstract}


\section{Introduction}

Extragalactic jets are amongst the most powerful phenomena in the universe and
observable up to high redshift \citep{MileyDeBreuck2008}. Jet activity is
generally accompanied by synchrotron emission from relativistic electrons in the
magnetized jet plasma, which is most prominently observable at radio
frequencies. Thus radio observations provided us with much insight into jet
morphology (beams, knots, hotspots, lobes/cocoons), classification into
low-power FR I and high-power FR II sources \citep{FanaroffRiley1974}, magnetic
field strengths estimated by minimum energy arguments in the range of a few to
several hundreds of microgauss in hotspots \citep{Bridle1982,Meisenheimer+1989}
as well as magnetic field topology from polarization measurements
\citep{BridlePerley1984} (high power jets generally show magnetic fields
parallel to the jet axis) and age estimates from spectral ageing
\citep{AlexanderLeahy1987,Carilli+1991} of a few $\times \, 10^7$ years.
Extragalactic jets show prominent cocoons, often only partly visible as lobes,
with total length to total width aspect ratios mostly between $3$ and $10$
\citep{Mullin+2008}. However, it must be cautioned that visible radio emission
does not necessarily trace all the bulk plasma, but depends on magnetic field
strength and electron acceleration.

The current X-ray observatories \emph{Chandra} and \emph{XMM-Newton} give rise to a
complementary view on jets, observing bremsstrahlung radiation from the thermal
ambient gas \citep{Smith+2002} and inverse-Compton emission from the jet cocoon
\citep{HardcastleCroston2005,Croston+2005}, the latter suggesting near-equipartition magnetic
fields. These emission processes are less dependent on microphysics and thus
easier to connect with theoretical results and furthermore contain information
about the history of these sources. 

An extreme example for this is MS0735.6+7421 \citep{McNamara+2005}, where radio
emission shows a weak source, while the spatially coincident X-ray cavities
reveal the true average power of the AGN ($1.7 \times 10^{46} \, \mathrm{erg\,
s}^{-1}$), which is a factor of $\sim \!\! 10^5$ higher.
The jet cocoon displaces the ambient thermal gas and drives a bow shock outward.
Both $pV$ work and the bow shock contain signatures of the jet properties and
may be excellent diagnostic tools. Nearly three dozen clusters were found to
have cavities \citep{McNamaraNulsen2007} and generally show weak bow shocks (Mach
1--2) with aspect ratios (length/width) not much above unity. 

Radio observations of high-power FR II jets show wide cocoons
\citep{Mullin+2008}, which are partly (and more completely at low frequencies)
visible as radio lobes. Both jet head and cocoon--ambient gas interface appear
smoother in radio maps (e.g. Cygnus A \citep{Lazio+2006}, Pictor A
\citep{Perley+1997}, but also Hercules A \citep{GizaniLeahy2003}) than they do
in hydrodynamic simulations; and while the emission of synchrotron radiation
obviously indicates the presence of magnetic fields, their importance for jet
dynamics and the contact surface is still unclear.

On the theoretical side, early numerical simulations of supersonic jets
\citep{Norman+1982} already exhibited basic structures seen in observations of
extragalactic radio sources (working surface, cocoon, bow shock) and showed that
pronounced cocoons are properties of jets with much lower density than the ambient
medium (light jets), although the slow propagation of the jet makes simulations
of these computationally very expensive. With the availability of more computing
power and new codes, simulations of the long-term evolution
\citep{Reynolds+2002}, in three dimensions
\citep{BalsaraNorman1992,Clarke+1997,KrauseVLJ2,Heinz+2006}, of very light
\citep{KrauseVLJ1,Saxton+2002HerA,Saxton+2002PicA,CarvalhoODea2002a,CarvalhoODea2002b,
Zanni+2003,SutherlandBicknell2007} and relativistic jets
\citep{Aloy+1999,Komissarov1999,RosenHughes+1999,Hardee2000,Leismann+2005}
became possible. Furthermore, effects of magnetic fields were examined
\citep{Clarke+1986,Lind+1989,Koessl+1990a,HardeeClarke1995a,
Tregillis+2001,Tregillis+2004,ONeill+2005,Li+2006,Mizuno+2007,Keppens+2008}, 
although only for relatively dense jets (density contrast $\ge 10^{-2}$). 
In this paper, we extend the studies of very light jets to include magnetic
fields.

To explore the interaction of jets with the ambient intra-cluster
medium and and the impact of magnetic fields, we performed a series of
magnetohydrodynamic (MHD) simulations of very light jets on the scale of up to
200 kpc (200 jet radii) with a globally consistent magnetic field configuration.
A constant ambient density was used to avoid effects of a declining cluster gas
density on the structural properties which could contaminate the effect of
magnetic fields, while the effects of a density profile previously were described in \citet{KrauseVLJ2} for the axisymmetric and threedimensional case.

After a description of our simulation setup and the numerical method, our results
are described: first about the morphology and dynamics, the evolution of the bow
shock and the cocoon; then entrainment of ambient gas and the energy budget; and
finally the magnetic fields and their evolution as well as their impact on
morphology and propagation. Results are then discussed and put in context with
observational findings.

To compensate for different propagation speeds of jets with differing density
contrasts, plots will use the axial bow shock diameter or jet length, where
appropriate. ``Full length'' refers to the whole simulated length (considering both
jets), while ``full width'' refers to twice the measured (radial) distance from
the axis.

\section{Setup and Numerical Method}

\subsection{Setup}

The idea behind the present study was to explore the behaviour of very light
jets with non-dominant magnetic fields in a cluster environment using a
plausible global setup for the plasma and the magnetic fields, but still keeping
the setup simple enough to see the working physical processes clearly, which is much
harder for a complex setup. We performed 2.5D simulations (axisymmetric with 3D
vector fields) of both purely hydrodynamic and MHD jets on the scale of up to
$200$ kpc with a constant ambient gas density, where density contrasts $\eta =
\rho_\text{j}/\rho_\text{a}$ were varied between $10^{-1}$ and $10^{-4}$ to see
its effects on the simulation. Jet speed, beam radius, sonic Mach number and
{magnitude of the helical magnetic field \citep{Gabuzda+2008}} were kept fixed, thus yielding a kinetic jet power
$L_\mathrm{kin} = \pi r_\jet^2 \eta \rho_\amb v_\jet^3$ and plasma $\beta = 8
\pi p / B^2$ varying with density contrast.  A summary of the parameters is
given in Tab.~\ref{tab:simpara-common} and \ref{tab:simpara-var}.  The
simulations are labelled by a letter and a numeral, indicating the inclusion of
magnetic fields (M) vs. pure hydrodynamics (H) as well as their density
contrast.  Both simplifications -- axisymmetry and density distribution -- were
relaxed in a previous hydrodynamic study \citep{KrauseVLJ2} and their influence
is addressed later in the discussion. The initial gas distribution was randomly
perturbed on the resolution scale to break symmetry between both jets.

The bipolar (back-to-back) jets were injected by a cylindrical nozzle (radius
$r_\jet$, length $2 r_\jet$) along the Z axis in cylindrical coordinates ($Z, R,
\phi$), hence allowing for interaction of the backflows in the midplane. The jet
radius is resolved by 20 cells. Fully ionized hydrogen ($\gamma=5/3$) was
assumed for both the jet and the external medium. A compressible tracer field
was advected with the flow, using a value of $1$ for the ambient gas, and $0$ 
and $-1$ for the jets, allowing to trace back the origin of the plasma. Optically
thin cooling is included in the code but was switched off, since the cooling
times for our setup ($\gg \! 10^8$ years) are significantly longer than the
simulated time-scale, even for the shocked ambient gas. This choice also makes
the simulations scalable, e.g. to other values of the jet radius, which was
chosen arbitrarily as $r_\jet = 1$ kpc.

In the jet nozzle, all hydrodynamic variables (density, pressure and velocity)
are kept constant at all times and a toroidal field $B_\phi \propto
\mathrm{sgn}(Z) \, \sin^4(\pi R/r_\jet) \, R/r_\jet$ is prescribed there, being
zero outside the nozzle. A dipolar field centred on the origin is used as
initial condition for the whole computational domain (magnetic moment aligned
with the jet axis) although it is mostly confined to the jet due to the strong
decrease in magnitude with distance.  For global simulations, the $\nabla \cdot
\vec{B}$ constraint enforces closed field lines, which is satisfied by a dipolar
field configuration, but not by the common setup of an infinite axial field,
which locally, but not globally fulfills the constraint. Thus, in our setup, the
magnetized jet plasma propagates into the essentially nonmagnetic ambient
matter. For M3 and especially M4, the magnetic fields become dynamically
important and influence the appearance, so another run with lowered magnetic
fields was performed in addition (M4L), which is more in line with the other
jets. These lightest jets are addressed more specifically in
Sect.~\ref{sec:lightestjet}. 
\begin{table}
  \caption{Common simulation setup parameters. Nozzle averages are restricted to
  $R \le 0.9 \, r_\mathrm{j}$ ($\le 0.8 \, r_\mathrm{j}$ for M4 and M4L).}
  \label{tab:simpara-common}
  \centering
  \begin{tabular}{l l l}
  \hline
  jet speed                 & $v_\jet$       & $0.6 \, c$                                                  \\
  jet sound speed           & $\cs$          & $0.1 \, c$                                                  \\
  jet radius                & $r_\jet$       & $1 \, \mathrm{kpc}$                                         \\
  ambient gas density       & $\rho_\amb$    & $10^{-2} \, m\mathrm{_p \, cm^{-3}}$                        \\
  ambient gas temperature   & $T_\amb$       & $5 \times 10^7 \, \mathrm{K}$                               \\
  jet nozzle magnetic field & $\avg{B_\pol}$ & $18.1 \, \mathrm{\umu G} \;$ (M4L: $1.81 \, \mathrm{\umu G}$) \\
                            & $\avg{B_\phi}$ & $7.5 \, \mathrm{\umu G}\;$ (M4L: $0.75 \, \mathrm{\umu G}$)   \\
  \hline
  \end{tabular}
\end{table}
\begin{table}
  \caption{Parameters for the different simulation runs. Nozzle averages are restricted to
  $R \le 0.9 \, r_\mathrm{j}$ ($\le 0.8 \, r_\mathrm{j}$ for M4 and M4L).
  $\beta$ values are typical over simulation run.}
  \label{tab:simpara-var}
  \centering
  \begin{tabular}{p{6ex} c r@{.}l c}
  \hline
  Run    & $\eta = \rho_\jet / \rho_\amb$ & \multicolumn{2}{c}{$\avg{\beta^{-1}}^{-1}$} &  $t_\mathrm{max}$ [Myr] \\
  \hline
  M1     & $10^{-1}$                      & 810                                         &        &   \phantom{0}6.7 \\
  M2     & $10^{-2}$                      & 81                                          &        &   10.9           \\
  M3     & $10^{-3}$                      & 8                                           & 1      &   16.5           \\
  M4     & $10^{-4}$                      & 0                                           & 89     &   47.5           \\
  M4L    & $10^{-4}$                      & 36                                          &        &   50.0           \\
  \hline
  \end{tabular}
\end{table}

The initial magnetic fields in Tab.~\ref{tab:simpara-common} are nozzle-averaged
initial values. As the poloidal field cannot be kept constant in the
nozzle without violating
$\nabla \cdot \vec{B} = 0$, this field can evolve with time due to the
interaction with the enclosing cocoon, quickly adjusting to $13 \,
\umu\mathrm{G}$ ($1.5 \, \umu\mathrm{G}$ for M4L) but then stays constant. For
these nozzle averages, only 90 per cent of the jet radius were considered for M1, M2,
M3 and 80 per cent for M4 and M4L to exclude cells at the shearing boundary of the
nozzle, where high magnetic fields and opposite field directions can occur,
while the core of the jet is unchanged. For plasma $\beta$, we use the
volume-averaged harmonic mean, as very weakly magnetized regions otherwise would
misguidingly dominate the average.

Since the jets are very underdense with respect to the ambient gas and have a
high internal sound speed, a reconfinement shock develops already very near the
jet nozzle. This shock establishes pressure balance between the jet beam and the
cocoon, resulting in a pressure-confined beam. In contrast to freely expanding
jets, any imposed opening angle becomes unimportant once the beam is reconfined
already near the jet inlet. Hence, in contrast to heavier jets, underdense jets quickly
find pressure balance with their environment.

The simulations were run until they reach the boundary of the uniform grid which has
$4000 \times 800$ or $4000 \times 1600$ cells (for M4 and M4L)
and the jet radius ($r_\jet$) is resolved with 20 cells.

In the following, we will focus only on the MHD jets, as their hydro
counterparts are only for comparison (set up exactly as the MHD jets with
vanishing magnetic field).

\subsection{Method}

The simulations were performed using the NIRVANA code \citep{ZieglerYorke1997}, which
numerically solves the nonrelativistic magnetohydrodynamic equations in
three dimensions in cartesian, cylindrical or spherical coordinates.  It is
based on a finite-differences discretization in an explicit formulation using
operator splitting and uses van Leer's interpolation scheme, which is second
order accurate. The advection part is solved in a conservative form and the
magnetic fields are evolved using the constrained transport method, which
conserves $\nabla\cdot\vec{B}$ up to machine roundoff errors. The code was
vectorized and shared-memory parallelized \citep{HLRS2006} for the
NEC SX-6 and SX-8. 

\section{Results}

\subsection{Morphology}

\begin{figure*}
  \centering
  \includegraphics[width=0.99\linewidth]{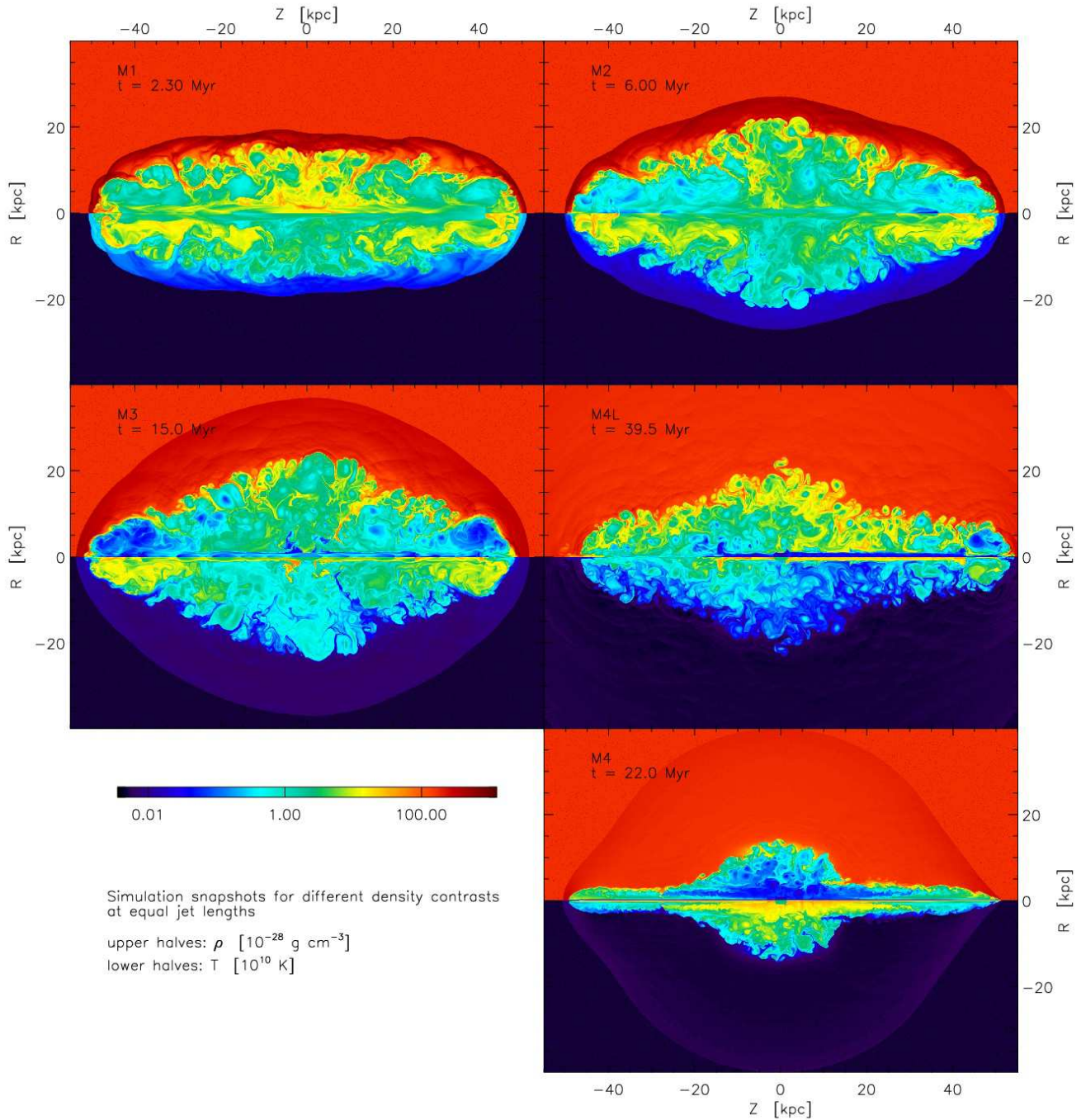}
  \caption{Density and temperature for all runs, each at a jet length of
  $\approx \! 100$ kpc. The upper halves of each panel show the logarithmic density
  in units of $10^{-28}$ g/cm$^3$, the lower halves show the logarithm of
  temperature in units of $10^{10}$ K. Panels additionally are labelled by their
  respective time.}
  \label{fig:eta-map}
\end{figure*}
The density and temperature maps of Fig.~\ref{fig:eta-map} show snapshots of all
runs at a full jet length of $100$ kpc, respectively. In the following, M3
mostly will be used for figures as it has the strongest non-dominant magnetic
fields, therefore showing effects of the magnetic fields best and allowing
for comparison of features between different figures. 

The jet backflow blows up a pronounced cocoon, surrounded by a thick shell of
shocked ambient matter. Dense ambient gas is mixed into the cocoon in
finger-like structures due to Kelvin--Helmholtz instabilities at the contact
surface. Near the jet heads, this instability is suppressed by the magnetic
field, which leads to a smoother appearance there. In purely hydrodynamic
simulations, this stabilization is absent. 

The cocoon is highly turbulent and vortices hitting the jet beam can easily
destabilize, deflect or disrupt it if jet densities are low. The Mach number
varies considerably along the beam 
\citep{KrauseVLJ1,Saxton+2002PicA} and there is
no stable ``Mach disc'' as seen for heavier jets -- the terminal shock moves
back and forth and often is not clearly defined.

Because very light jets only propagate slowly, basically hitting the ambient gas
as a ``solid wall'', the backflow is strong and the turbulence makes the
interaction between both jets in the midplane important. Such jets have to be
simulated bipolarly to describe the lateral expansion and hence the global
appearance correctly. If only one jet were simulated, for very light jets the
result would strongly depend on the boundary condition in the equatorial plane
\citep{Saxton+2002PicA}.

The surrounding ambient gas is pushed outwards by the cocoon pressure, driving a
bow shock outwards. The bow shock for very light jets is different in its shape
and strength from that of heavier jets (see Sect.~\ref{sec:bscoevol}). It is
additionally changed by a density profile in the external medium
\citep{KrauseVLJ2}, which increases the aspect ratio with time because $\eta$
increases at the jet head and thus shows cylindrical cocoons.

\subsection{Defining the cocoon}
\label{sec:cocoondefinition}

In the following, we not only measure properties of the bow shock, which is easy
to pin down, but also of the cocoon. While generally we define the cocoon as the
region, which is filled by jet-originated matter (not including the beam itself)
this definition has to be made in more detail for the simulation analysis. 
The strong backflow and the fragile beams of very light jets make the
distinction between cocoon and beam difficult, while mixing at the contact
discontinuity complicates the assignment of cell to cocoon or ambient matter. 
While we do not attempt to distinguish between beam and cocoon if not stated
explicitly (it only seems necessary for energetic investigations), the
distinction between cocoon and ambient matter is necessary especially for the
entrainment measurements later and thus is described in more detail in this
subsection, along with the measurement of the cocoon properties. 

\subsubsection{Cell assignment}

Two properties can be used for the distinction between cocoon and ambient
matter: the (compressible) tracer field and the toroidal magnetic field.  Tracer
field values of $1$ and above indicate undisturbed and shocked ambient matter.
This is available in all simulations, but mixing with jet matter at the contact
discontinuity (due to finite resolution) lowers the tracer and thus requires a
threshold value. The cocoon mass is especially sensitive to this threshold
value, as the density of the ambient gas is much higher and thus causes large
changes of the cocoon mass if the border is shifted.
\begin{figure}
  \centering
  \includegraphics[width=0.95\linewidth]{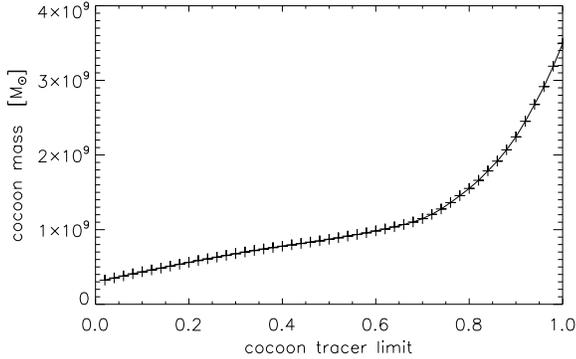}
  \caption{Cocoon mass depending on tracer field threshold (mass of all cells
  with tracer below the threshold), for simulation M3 at $t = 15$ Myr.}
  \label{fig:Mco-trlimit}
\end{figure}
Figure~\ref{fig:Mco-trlimit} shows the cocoon mass for a range of tracer
thresholds. The injected mass at this time is only $4 \times 10^6 \, M_\odot$; 
measured mass above this value is the entrained ambient gas mass. 

\begin{figure}
  \centering
  \includegraphics[width=0.95\linewidth]{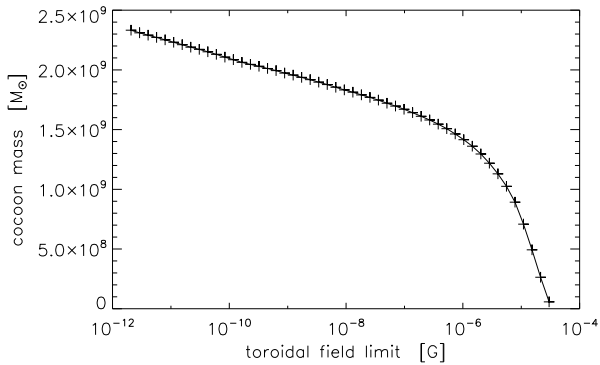}
  \caption{Cocoon mass depending on toroidal field threshold (mass of all cells
  with toroidal field magnitude above the threshold). Simulation M3 at $t = 15$ Myr.}
  \label{fig:Mco-maglimit}
\end{figure}
In contrast, the toroidal field strength can be used for separation, as the
toroidal field is zero initially in the ambient medium and is conserved
independent from the other field components.
Figure \ref{fig:Mco-maglimit} shows the cocoon mass depending on the toroidal
magnetic field threshold. Using this method, even cells with only a small mass
fraction of jet matter can be assigned to the cocoon. There is a clear break
visible, but the cocoon mass continuously increases for lower threshold values
until machine accuracy is reached. This has two major problems: First, it 
naturally is not available for the pure hydro simulations and thus cannot be used
to compare HD with MHD simulations. Second, the high sensitivity to jet matter
is not a real advantage, as the mass values do not nicely converge and we have
to choose a threshold.

In the following we will use a tracer threshold of $0.5$ which is available for
HD and MHD models and gives cocoon masses that do not strongly depend on the
tracer threshold. Furthermore, it selects the regions one would consider cocoon
also by looking at the other physical variables.

\subsubsection{Shape measurement}

\begin{figure}
  \centering
  \includegraphics[width=0.95\linewidth]{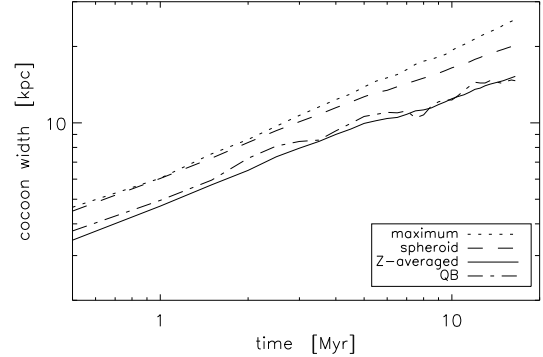}
  \caption{Evolution of the cocoon width for different width definitions for 
  simulation M3. A tracer limit of 0.5 was used for cell assignment.}
  \label{fig:co-widthdef}
\end{figure}

To characterize the width of the cocoon, we checked four different measures,
which will generally give different results due to the ragged shape of the
contact surface. Widths are measured from the symmetry axis ($R=0$, jet
channel) and thus are only ``half widths''.
Figure~\ref{fig:co-widthdef} shows the temporal evolution of the cocoon width
definitions
\begin{enumerate}
  \item maximum width: measured at the maximum $R$ position of a cocoon cell,
  \item average width: $Z$-averaged over the full jet length,
  \item QB width: measured at one quarter the full jet length backwards of the
    jet heads,
  \item spheroid width: semi-minor axis of a spheroid with a volume equal to the cocoon
    volume and the semi-major axis equal to half the full jet length.
\end{enumerate}
\begin{figure}
  \centering
  \includegraphics[width=0.95\linewidth]{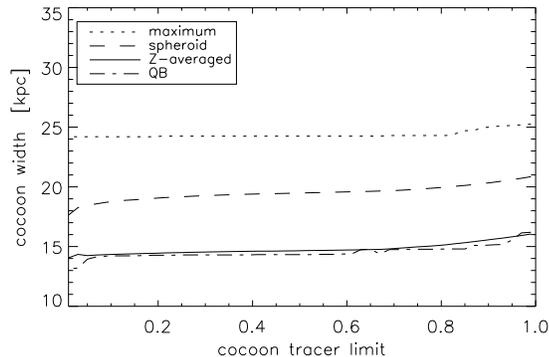}
  \caption{Cocoon width for different width definitions as function of the used
  tracer limit. For simulation M3 at 15 Myr.}
  \label{fig:cowidth-tr}
\end{figure}

The QB width is clearly much dependent on vortices near the contact
surface and not very straight. Despite that, it grows similar to the Z-averaged
width, which mostly has the lowest width value of all four definitions. The spheroid width
lies between the maximum width and the Z-averaged width. All of theses measures can
be approximated by power laws, although with somewhat different parameters.

In contrast to the cocoon mass, the cocoon shape does not depend strongly on the
tracer limit that is used for its determination (Fig.~\ref{fig:cowidth-tr}). For
limits around 0.5, the differences between width definitions is larger than the
dependence on the tracer limit.

\subsection{Evolution of bow shock and cocoon}
\label{sec:bscoevol}

\subsubsection{Cocoon pressure evolution}

\begin{figure}
  \centering
  \includegraphics[width=0.95\linewidth]{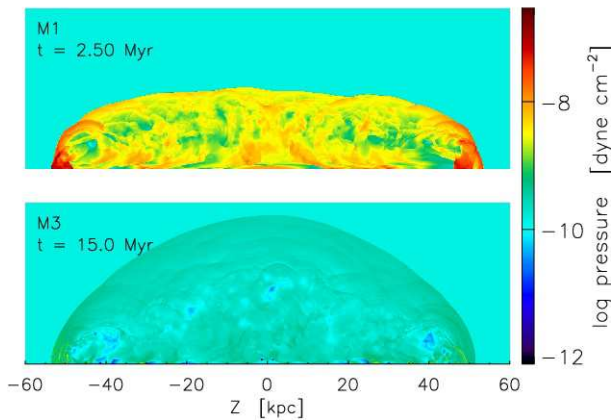}
  \caption{Pressure maps in logarithmic scaling for simulation M1 and M3 at equal lengths.}
  \label{fig:etacomp-pressure}
\end{figure}
The low jet density has two main consequences for the evolution of the cocoon
pressure: one is the lower jet power (for a fixed jet bulk velocity), which
results in a lower cocoon pressure and a generally weaker bow shock. The other
is the slow jet head propagation, which lowers the propagation time-scale
compared to the dynamical time-scale of travelling pressure waves within the
cocoon. Pressure waves from the jet head together with waves induced by
turbulent motion and mixing in the cocoon, try to establish pressure balance
within the cocoon and between cocoon and ambient gas, driving the lateral
expansion of the cocoon.

Figure~\ref{fig:etacomp-pressure} shows pressure maps of jets with $\eta=10^{-1}$
and $\eta=10^{-3}$ at the same lengths.  The cocoon of the heavier jet is
overpressured by a factor of $20$ with respect to the ambient gas, while being a
factor of only $1.5$ for the lighter jet (and $4.9$ for this jet at the time of
the M1 image).

The strong evolution towards pressure balance is responsible for the much less
pronounced high-pressure regions between Mach disc and the advancing bow shock.
The bow shock has an elliptical shape with less directional dependence of its
strength, more similar to an overpressured bubble, although it is still stronger
in axial direction (see Sect.~\ref{sec:bowshock}). 

\begin{figure}
  \centering
  \includegraphics[width=0.95\linewidth]{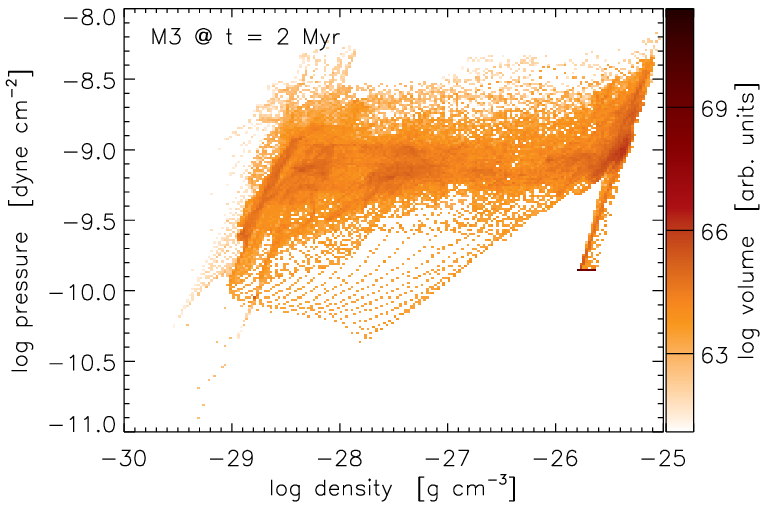}
  \includegraphics[width=0.95\linewidth]{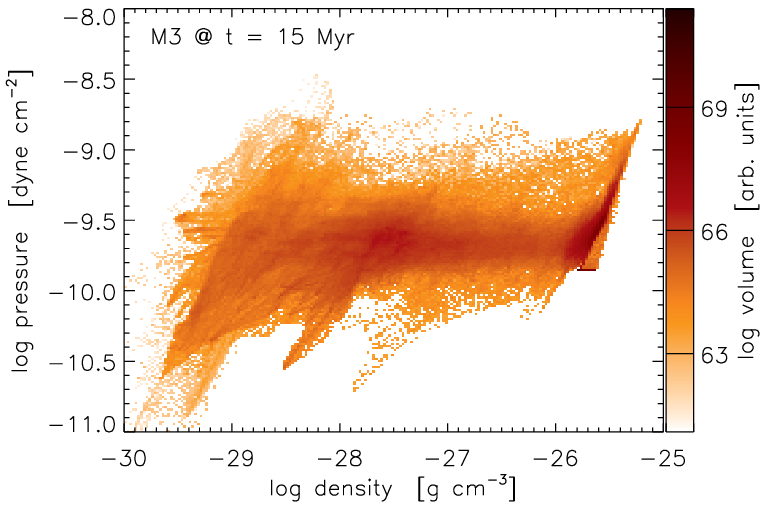}
  \caption{Pressure vs.\ density histogram for the M3 jet after
  $2$ and $15$ Myr. Volume-weighted counts, pressure and density are shown
  logarithmically.}
  \label{fig:pndiag}
\end{figure}
The quick pressure adjustment can also be seen in the pressure--density diagrams
of Fig.~\ref{fig:pndiag}. The ambient gas is described by the patch near $(-26,
-10)$, the jet nozzle by the cells around $(-29,-10$). Adiabatic compression and
expansion leads to the oblique and longish features present at different
positions. Top right of the jet nozzle position are the cocoon grid points,
which spread over a large range of density to the right because of mixing with
shocked ambient gas, which is the elongated feature top right of the ambient gas
position. Comparing the two different simulation snapshots, we find that the
pressure distribution is quickly adjusting towards the external pressure, in
agreement to the findings in \citet{KrauseVLJ1}, and the cocoon is not strongly
overpressured anymore.

\begin{figure}
  \centering
  \includegraphics[width=0.95\linewidth]{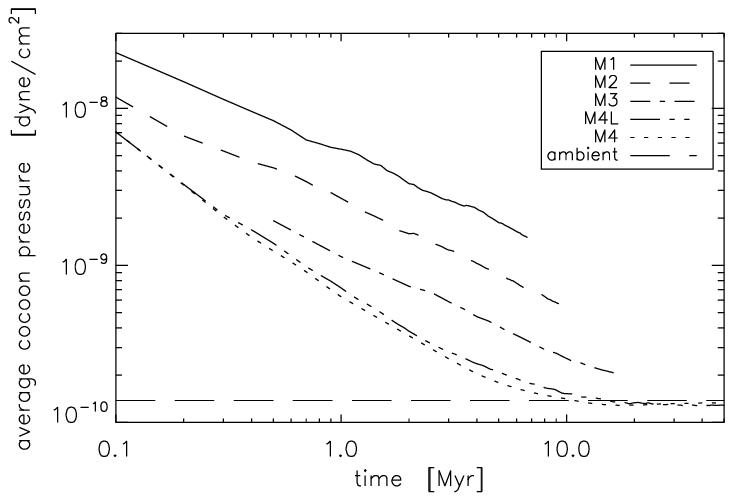}
  \includegraphics[width=0.95\linewidth]{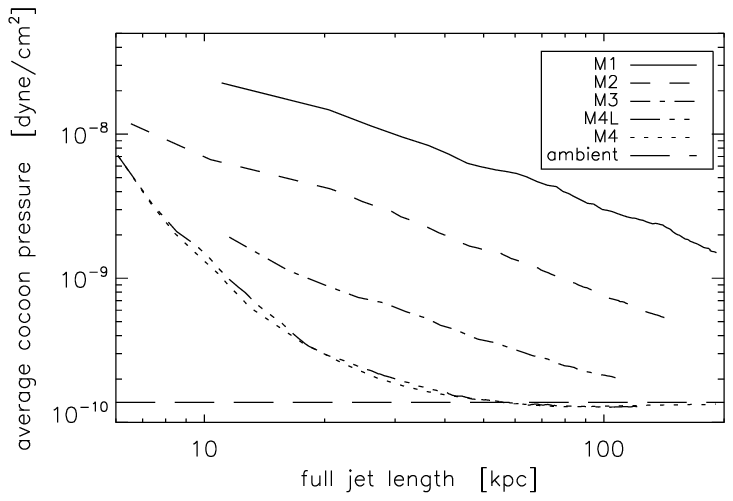}
  \caption{Evolution of the volume-averaged cocoon pressure as function of time
  or the jet length for the different models. The cocoon is defined by a tracer
  limit of $0.5$. MHD models are shown in thick lines, corresponding hydro
  models in thin lines, the ambient pressure is indicated by the long-dashed
  line for comparison.}
  \label{fig:copressure-len}
\end{figure}
Another view on this is the average cocoon pressure, shown in
Fig.~\ref{fig:copressure-len}, which has a power-law-like behaviour. For the three
models M1, M2 and M3, it strikingly decreases with the reciprocal jet length
(Tab.~\ref{tab:bscoevol}). While M1 at the end of the simulation is still very
overpressured, the cocoon pressure of M3 is already near the ambient gas
pressure. 

M4 and M4L seem to deviate from this behaviour. At the beginning this is mainly
a consequence of the longer-lasting relaxation from initial conditions, where strong shocks are
thermalized efficiently, increasing the pressure in the early ``cocoon bubble''
and because the early phase is shown with higher time resolution.
After a jet length of $20$ kpc has been reached, they fit into the
behaviour of the other simulations, but, as they soon reach the ambient
pressure, settle to its value. 

It is clear that the cocoon pressure cannot drop much below the ambient
pressure and thus approaches its value. At this point we expect the bow shock to
softly turn into an ordinary sound wave. This is just about to happen in the last
snapshots of M4 and M4L, where there is only a very weak density jump,
corresponding to Mach 1.05. The exact value of the average cocoon pressure is
insensitive to the exact definition of the cocoon (see
Sect.~\ref{sec:cocoondefinition}), but can drop slightly below the ambient
pressure due to pressure variation within the cocoon (which can still be as
strong as a factor of 2).

\begin{figure}
  \centering
  \includegraphics[width=0.95\linewidth]{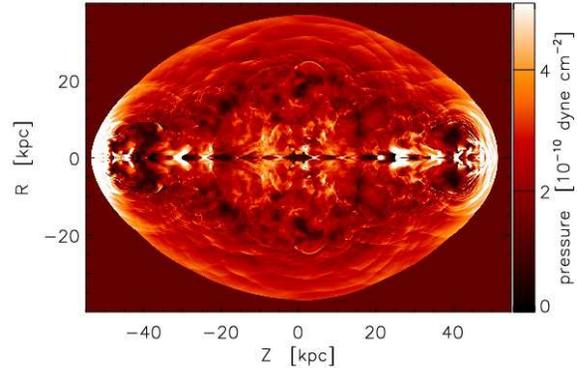}
  \caption{Linearly scaled pressure map of M3 at 15 Myr. Values above $5\times
  10^{-10}$ dyne cm$^{-2}$ are clipped.}
  \label{fig:sag-waves}
\end{figure}

The past bow shock is not the only sound wave testifying to the expanding
cocoon. Already much before the shock decays, waves and ripples can be seen in
the shocked ambient gas (Fig.~\ref{fig:sag-waves}).

\subsubsection{Bow Shock}
\label{sec:bowshock}

\begin{table*}
  \caption{Power law exponents for the bow shock and cocoon evolution. 
  The exponents $p$ in the table are fits to power laws as function of time
  (for the lengths and widths) or as function of full jet length (for the
  average cocoon pressure), considering only the data points which match the
  criterion in the third line. A minus (-) denotes that no reasonable power law
  fit could be done.
  }
  \label{tab:bscoevol}
  \centering
  \begin{tabular}{l c c c c c}
    \hline
    model  & bow shock full length & bow shock width & cocoon full length & cocoon average width & average cocoon pressure \\
           & $\propto t^p$         & $\propto t^p$   & $\propto t^p$      & $\propto t^p$        & $\propto l^p$           \\
           & $\ge$ 30 kpc          & $\ge$ 15 kpc    & $\ge$ 30 kpc       & $\ge$ 5 kpc          &                         \\
    \hline
    M1     & 0.63                  & 0.64            & 0.64               & 0.59                 & -0.99                   \\
    M2     & 0.63                  & 0.61            & 0.64               & 0.52                 & -1.04                   \\
    M3     & 0.68                  & 0.65            & 0.71               & 0.39                 & -0.95                   \\
    M4L    & 0.69                  & 0.71            & 0.64               & 0.24                 & -                       \\
    M4     & 0.80                  & 0.71            & 0.82               & -                    & -                       \\
    \hline
  \end{tabular}
\end{table*}

\begin{figure}
  \centering
  \includegraphics[width=0.95\linewidth]{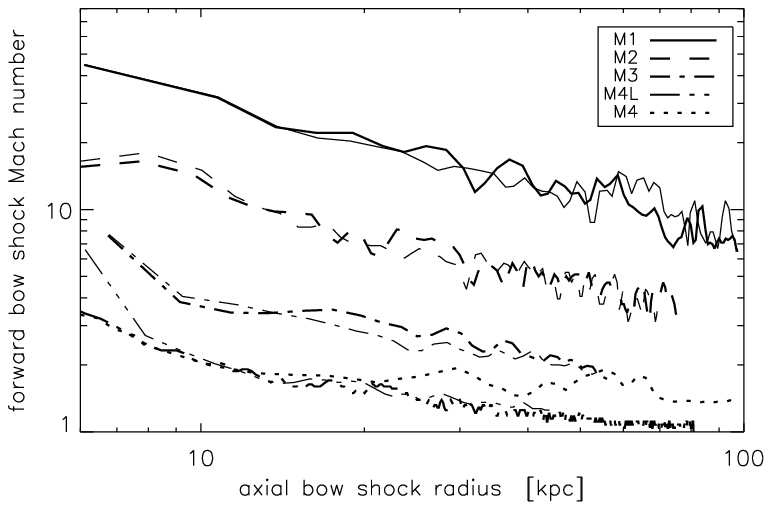}
  \includegraphics[width=0.95\linewidth]{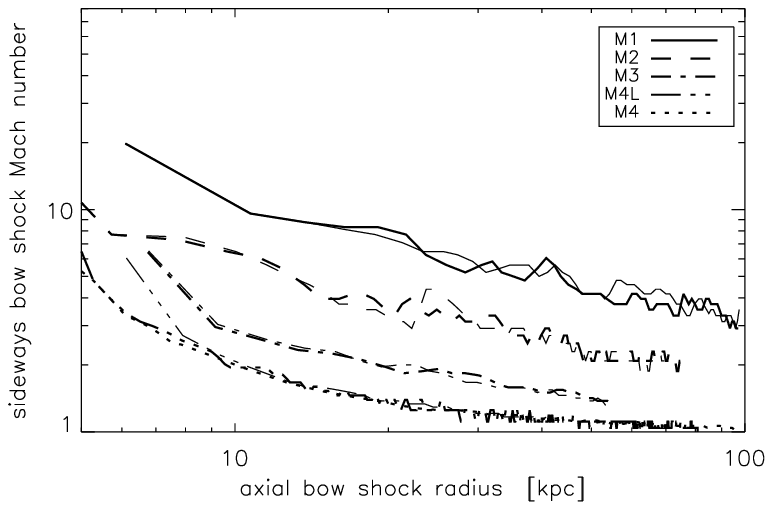}
  \caption{Evolution of the forward (top panel) and sideways (bottom panel) bow shock
  strength as a function of the monotonically increasing axial bow shock radius.
  Thick lines are MHD models, thin lines the corresponding hydro models for
  comparison.}
  \label{fig:bsstrengthevol}
\end{figure}
The quick decrease in cocoon pressure naturally affects the strength of the bow
shock as it is this pressure that drives the shock laterally.
Figure~\ref{fig:bsstrengthevol} shows the temporal evolution of the bow shock
strength, in terms of external Mach numbers, for the forward direction
(at $R=0$) as well as the lateral direction (at $Z=0$) for jets with different density
contrasts. 

\begin{figure}
  \centering
  \includegraphics[width=0.95\linewidth]{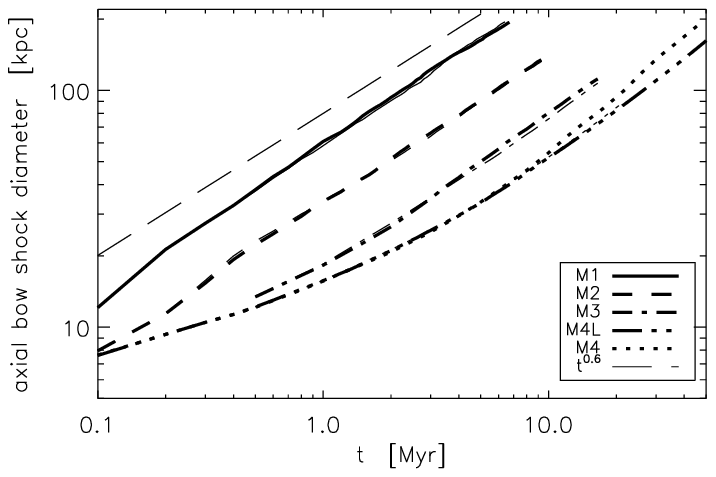}
  \includegraphics[width=0.95\linewidth]{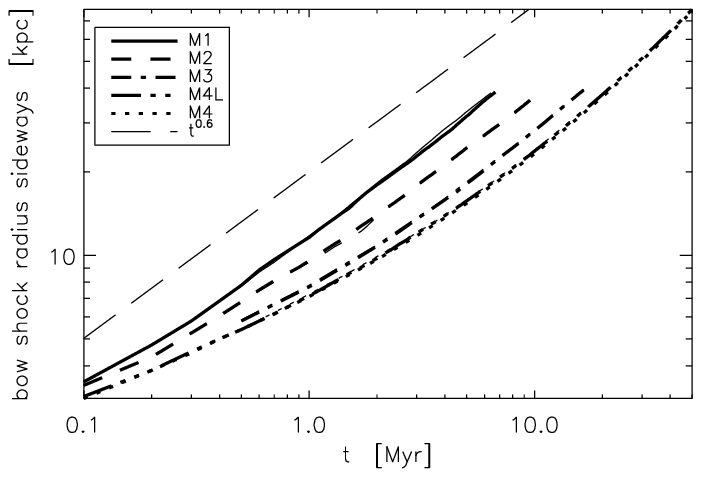}
  \caption{Propagation of the bow shock in axial (top panel) and lateral
  direction (bottom panel) for the simulated models as function of time. Thick
  lines are MHD models, thin lines the corresponding hydrodynamic models. A
  power law $\propto t^{0.6}$ is shown for comparison as long-dashed line.}
  \label{fig:bslenwidth}
\end{figure}
The bow shocks in forward direction are always stronger than the
sideways shocks due to the direct impact of the jet on to the ambient gas. The
lighter jets have a much weaker bow shock in all directions and the differences
between the axial and lateral direction are much less pronounced. 

The axial diameter of the bow shock grows as a power law with exponents
$\approx 0.65$ (Fig.~\ref{fig:bslenwidth} and Table~\ref{tab:bscoevol}). For the
lateral propagation we find similar exponents. This behaviour agrees with
self-similar jet models
\citep{Falle1991,Begelman1996,KaiserAlexander1997,KomissarovFalle1998} and 
the spherical blastwave approximation \citep{KrauseVLJ1}, which predict an
exponent of $0.6$.
At very early times and lasting longer for the lighter jets, we find lower exponents,
as for a Sedov blast wave ($l \propto t^{0.4}$) from the initial conditions.

The lighter jets have generally lower Mach numbers, as their kinetic power is
lower, thus showing smaller bow shock velocities. The aforementioned analytical
models yield an expansion speed
\begin{equation}
  v(r) 
  = k \, \left(\frac{L_\mathrm{j}}{\pi \, \rho_\mathrm{a}}\right)^{1/3} \, r^{-2/3} 
  = k \, v_\jet \, \eta^{1/3} \, \left(\frac{r}{r_\jet}\right)^{-2/3},
  \label{eqn:bsspeed-r}
\end{equation}
that directly translates into the bow shock Mach number and describes the
scaling behaviour of the simulations reasonably well ($L_\jet$: jet power). We
find values for $k$ between $1.5$ and $2$ in axial direction and between $0.5$
and $2$ (M4L) in lateral direction, the latter being increasingly higher for
lighter jets.

A clear deviation from this behaviour is M4 at $t > 10$ Myr in axial direction.
The bow shock propagates much faster due to the formation of a nose cone.
The strong toroidal magnetic field collimates the jet, suppresses the
pronounced backflow of M4L, and the Lorentz force of the radial current gives
the jet additional thrust for the propagation (see \ref{sec:lightestjet}). The
other light jets (M3 and M4L) also may propagate somewhat faster due to
their appreciable magnetic fields.

\subsubsection{Cocoon}
\label{sec:cocoon}

\begin{figure}
  \centering
  \includegraphics[width=0.95\linewidth]{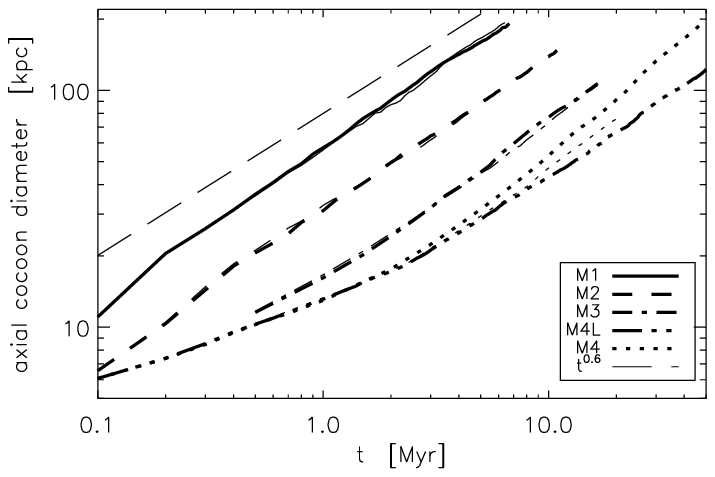}
  \includegraphics[width=0.95\linewidth]{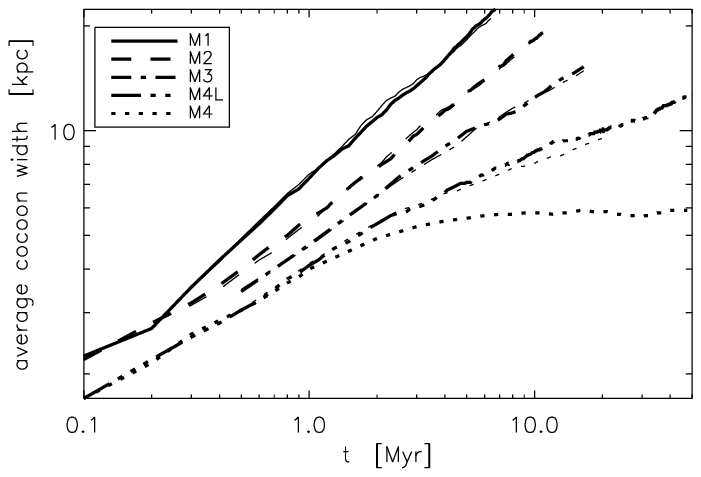}
  \caption{Time evolution of the cocoon: full length and average width as
  function of time. MHD models in thick lines, corresponding hydro models in
  thin lines.}
  \label{fig:colenwidth}
\end{figure}
As the jet pushes the bow shock forward in axial direction, the cocoon length
grows similar to the bow shock length (Fig.~\ref{fig:colenwidth}), showing a
power law behaviour with similar exponents. Again, M4 shows a higher exponent 
($0.82$) due to its additional thrust support in the nose cone. There might also 
be a slightly faster propagation for the M3 jet, where the magnetic
field is not too much below equipartition at the jet inlet (see
Fig.~\ref{tab:simpara-var}), although this might also be just a temporal effect
due to the jet--cocoon vortex interaction.

\begin{figure}
  \centering
  \includegraphics[width=0.95\linewidth]{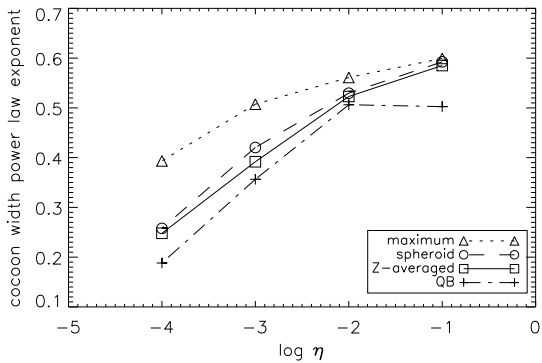}
  \caption{Power law exponents for cocoon widths as function of density
  contrast. Fits considered only data points where width $\ge 5$ kpc.}
  \label{fig:bscowidthexp}
\end{figure}

The cocoon width, in contrast, shows different power law exponents
depending on the density contrast, after the start-up phase 
is over. We find exponents of $0.59$ (M1), $0.52$ (M2), $0.39$ (M3) and $0.24$
(M4L) for the different models (Table~\ref{tab:bscoevol}). Thus there seems to be a clear trend of
decreasing exponents for lower jet densities, which holds true for all our
cocoon width measures (Fig.~\ref{fig:bscowidthexp}). Widths approaching an asymptotic
value might mimic a similar behaviour, but so far, this is beyond our simulation
data (except for M4). It seems reasonable that this is due to less
overpressured cocoons for lighter jets, as it is the cocoon pressure that drives
the lateral cocoon expansion \citep{KaiserAlexander1997,CarvalhoODea2002b}. If
the cocoon pressure equals the ambient pressure, the sideways expansion of the
cocoon would come to an end. 

\begin{figure}
  \centering
  \includegraphics[width=0.95\linewidth]{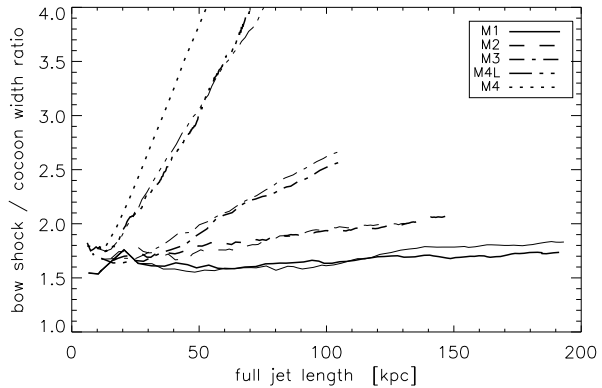}
  \caption{Bow shock / cocoon width ratio over time for the different
  simulations as function of the full jet length. MHD models in thick lines,
  corresponding hydro models in thin lines.}
  \label{fig:bscowidthratio}
\end{figure}
Another consequence of this is the lateral bow shocks, compared to
the corresponding cocoon width, being much further away for light jets 
(Fig.~\ref{fig:bscowidthratio}), as found by \citet{Zanni+2003}, too. Hence,
except for the H1/M1 models, the thick layer of shocked ambient gas grows
continuously. 

The expansion of the cocoon for M4 is much different. After the initial phase
the cocoon width settles down to a constant value and does not grow anymore. 
This is a consequence of the suppressed backflow in the nose cone,
which then cannot inflate the cocoon anymore.

\begin{figure}
  \centering
  \includegraphics[width=0.95\linewidth]{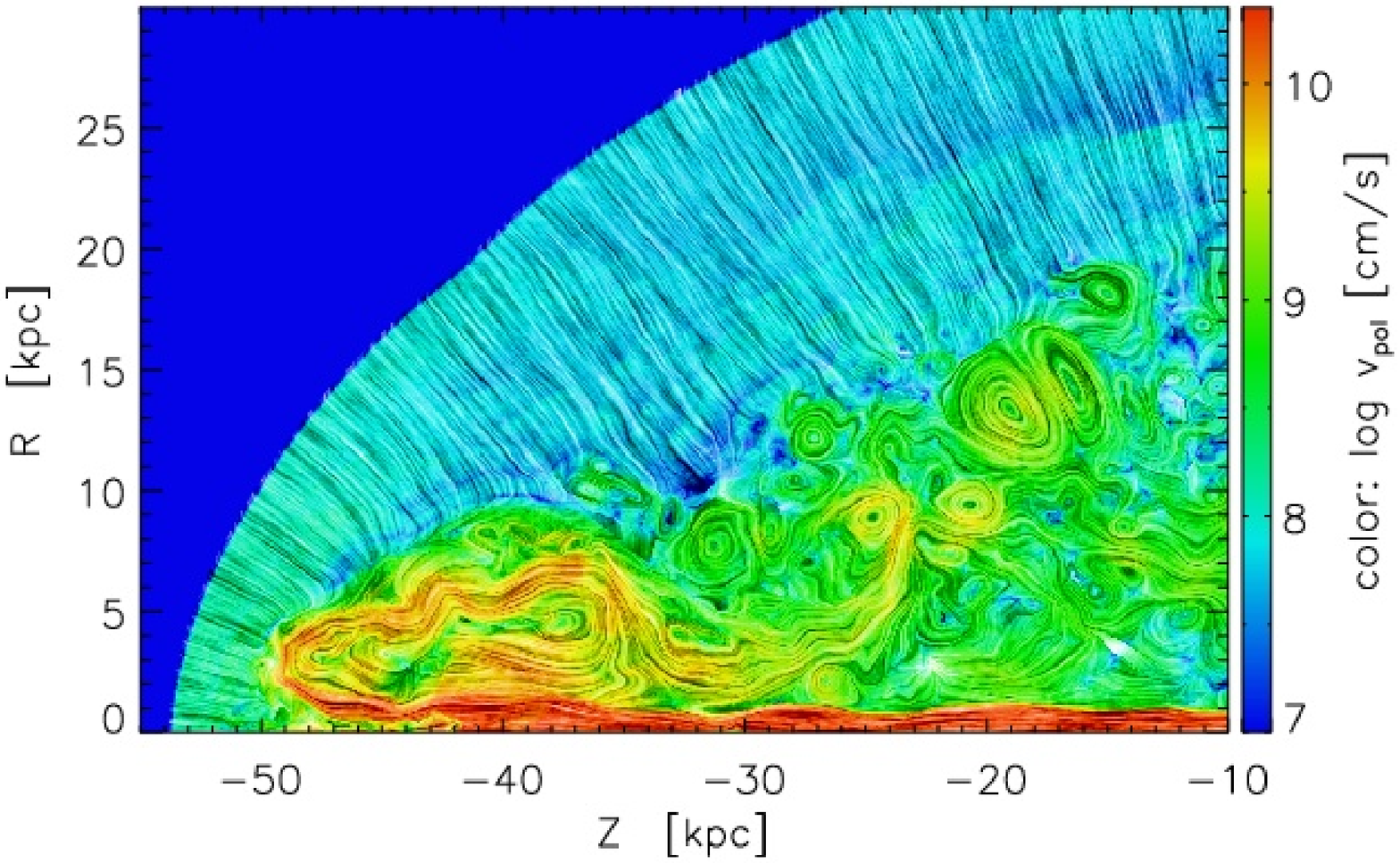}
  \includegraphics[width=0.95\linewidth]{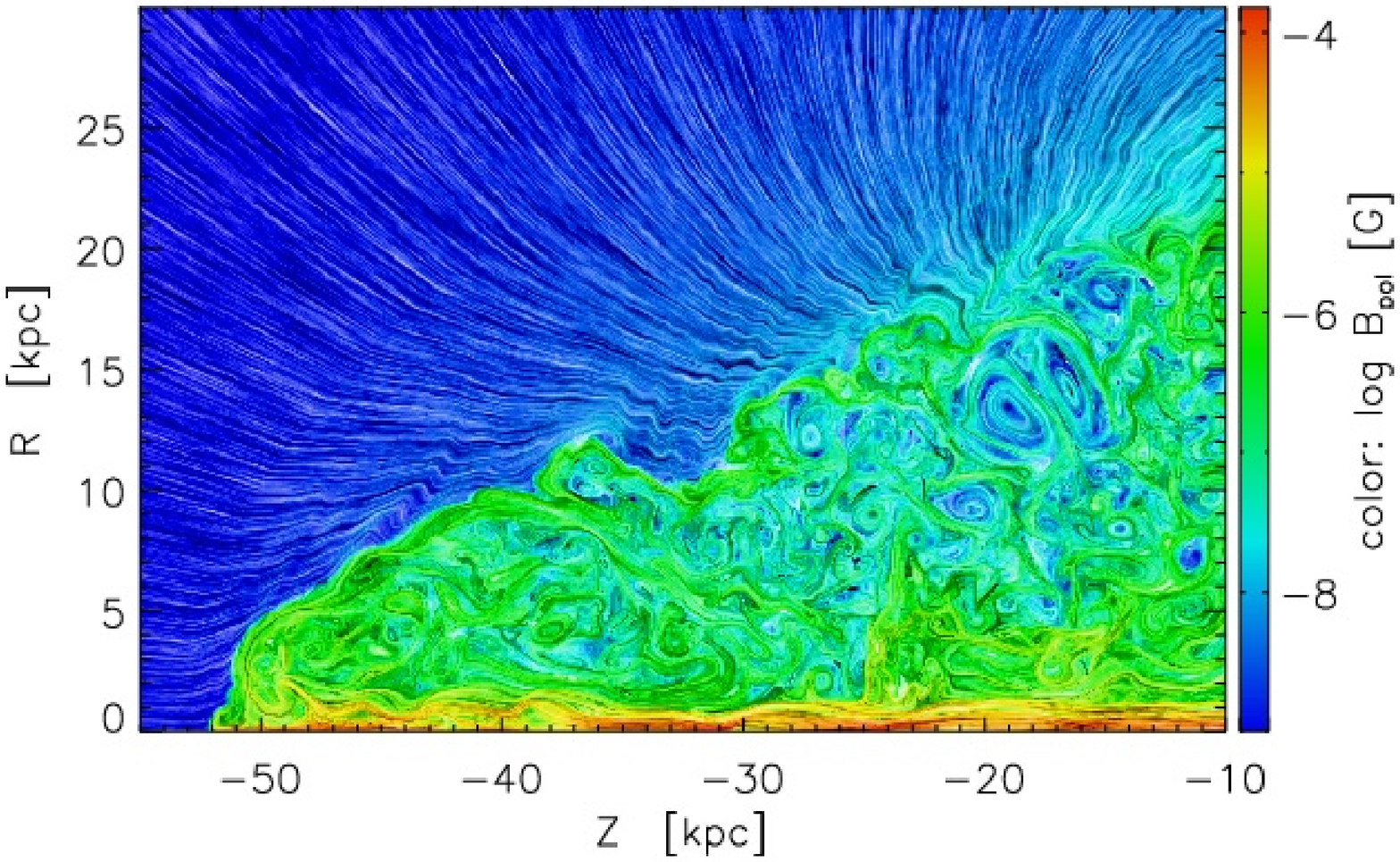}
  \caption{Velocity field (upper panel) and poloidal magnetic field (lower
  panel) around the jet head, displayed in LIC (line integral convolution)
  representation to show the small-scale vector field structure. The colours show
  the vector field magnitude, while the brightness modulation shows the field
  lines.}
  \label{fig:lic-maps}
\end{figure}

All simulations with non-dominant magnetic fields show pronounced turbulence in
their cocoons. This is evident from Fig.~\ref{fig:lic-maps}, which shows the
vector fields of velocity and poloidal magnetic fields in LIC (line integral
convolution) representation. The LIC technique \citep{CabralLeedom1993} allows the
fine-grained depiction of vector fields, especially suitable for turbulence,
where structures even on smallest scales are present due to the turbulent
cascade. We extended this to show the field magnitude, additionally, decomposing
the information into brightness (showing the field direction as stream lines)
and colour (field magnitude) in HLS colour space.

Cocoon turbulence is driven by quasi-periodic ``vortex shedding''
\citep{Norman+1982} in the jet head, which injects vortices into the the cocoon.
As these vortices move around and interact, vortex shedding affects the whole
cocoon and drives its turbulence. While it occurs in our heavier jets, too,
narrow cocoons suppress vortex interaction and the establishment of turbulence.
We note that there may be feedback on the driving mechanism, as cocoon vortices
perturb the jet beam and thus influence the vortex shedding process itself.

\subsubsection{Aspect Ratio}
\label{sec:aspectratio}

\begin{figure}
  \centering
  \includegraphics[width=0.95\linewidth]{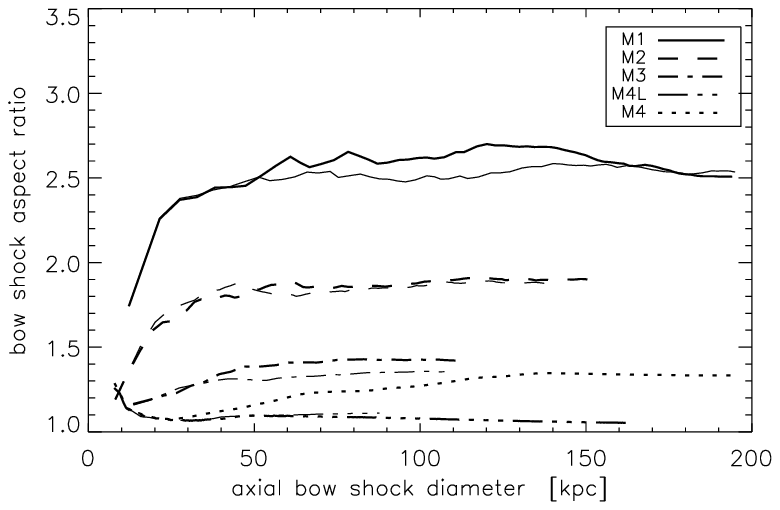}
  \includegraphics[width=0.95\linewidth]{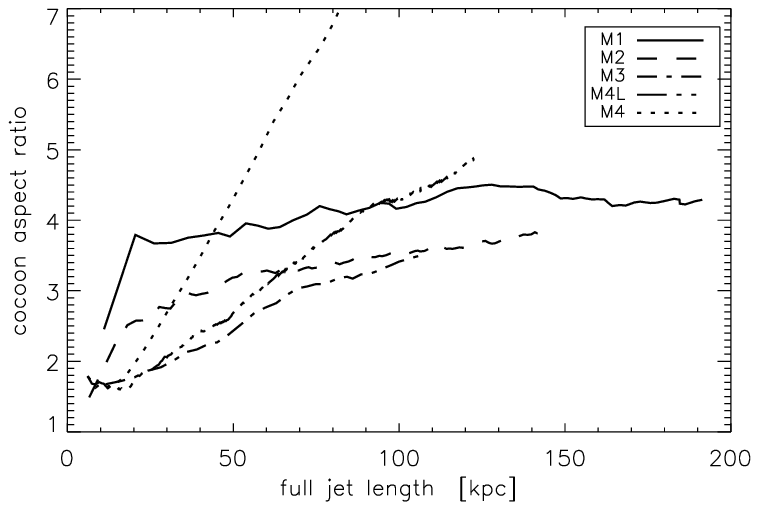}
  \caption{Aspect ratios $\aspectratio$ of bow shock (upper panel) and cocoon
  (lower panel) over full length. For the cocoon aspect ratios, the hydro models
  are omitted for clarity.} 
  \label{fig:aspectoverlength}
\end{figure}
A characteristic property of the bow shock or cocoon is their aspect ratio
$\aspectratio = \textrm{length} / \textrm{width}$
(Fig.~\ref{fig:aspectoverlength}).  Dependent on the density contrast, after a
short initial phase of spherical expansion ($\aspectratio \approx 1$), the bow
shock aspect ratios grow but converge for large bow shock diameters, approaching
$1$ for lighter jets ($\aspectratio = 1.4$ for M3 and $1.1$ for M4L). This means, the bow shock
approaches a spherical shape for very light jets. Once again, M4 is different,
as the propagation in axial direction is faster, yielding significantly higher
aspect ratios then M4L. 

The aspect ratios for the cocoons generally increase with jet length and 
are at early times systematically lower for the lighter jets. However, the
light jets soon increase their aspect ratio (earlier for lighter jets) and then
at later times, show aspect ratios even higher than their heavy counterparts. As
for the cocoon width evolution, we argue that this is due to cocoons, which
come to pressure balance with the ambient gas earlier, so that lateral cocoon
expansion stalls, but the axial propagation is still growing self-similarly.
By comparing Fig.~\ref{fig:copressure-len} and Fig.~\ref{fig:aspectoverlength}
one sees that once a source approaches pressure balance with its environment,
it drops out of self-similarity and increases its cocoon aspect ratio. For M3
this happens already early, while it does not happen for M1 until the end of the
simulation.

\subsection{Entrainment}

The jet backflow at the contact surface between the cocoon and the ambient gas
makes it Kelvin--Helmholtz unstable, and thus creates fingers of dense ambient
matter that reach into the cocoon and are entrained. In numerical simulations
this entrained gas is additionally mixed with the jet plasma due to finite
numerical resolution. The amount of entrainment can be measured in terms of the
cocoon mass since the mass of jet plasma usually is small compared to the
measured cocoon mass. Although the exact numbers depend on the exact cocoon
measurement definition (Sect.~\ref{sec:cocoondefinition}), this seems to be a
reasonably robust method.

\begin{figure}
  \centering
  \includegraphics[width=0.95\linewidth]{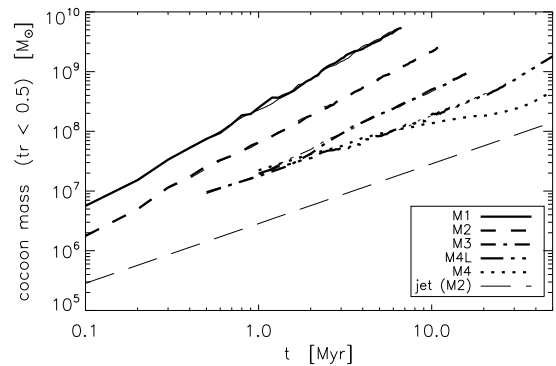}
  \caption{Evolution of the cocoon mass as measure for the entrainment of
  dense ambient gas. The long-dashed line gives the injected jet mass $M_\jet$
  for the M2 run for comparison ($M_\jet \propto \eta \, t$). MHD models as
  thick lines, corresponding hydro models as thin lines.}
  \label{fig:comass-t}
\end{figure}
\begin{figure}
  \centering
  \includegraphics[width=0.95\linewidth]{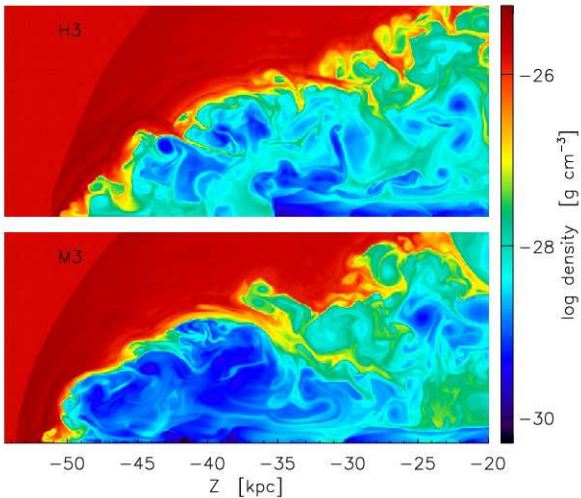}
  \caption{Comparison of hydro (H3) and MHD (M3) simulations at 15 Myr. The jet
  head region is much more pronounced, Kelvin--Helmholtz instabilities are
  damped, and the jet has propagated a bit further.}
  \label{fig:hydroMHDcomp}
\end{figure}
Figure~\ref{fig:comass-t} shows the time evolution of the cocoon mass.
The entrained mass grows with a power law exponent only slightly below the
exponent of the cocoon volume, showing a slowly decreasing but roughly constant
fraction (5--10 per cent) of the initial mass in the occupied volume. However, there is
no difference visible between purely hydrodynamic and MHD simulations in the
entrained mass, as would have been expected. The reason for this is the missing
stabilization of the contact surface, which is discussed later.
However, it is evident from M3 in Fig.~\ref{fig:hydroMHDcomp} that the
entrainment in the jet head is significantly smaller: the mass in a cylindrical
volume ($Z \in [-45, -35]$ kpc, radius $4$ kpc) in the head region of M3 is
$3 \times 10^5 \, M_\odot$ without magnetic fields (H3), compared to $9 \times
10^4 \, M_\odot$ in the magnetized case, which is more than a factor of $3$
lower. Hence entrainment is significantly suppressed in the jet head, but no
change could be measured regarding the whole cocoon volume.

\subsection{Energy budget}

From the quick balancing of pressure within the cocoon, one might expect
a strong conversion of (kinetic) jet power to thermal energy. This, in fact,
is measured for our simulations. 

\begin{figure}
  \centering
  \includegraphics[width=0.95\linewidth]{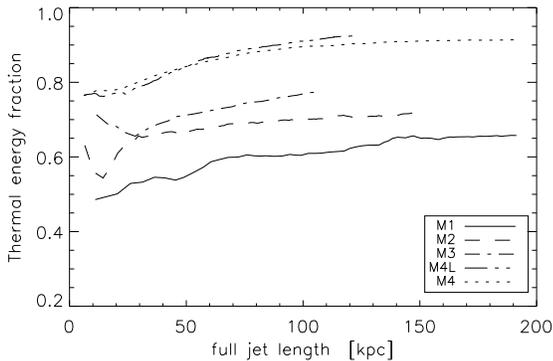}
  \caption{Gains in thermal energy as fraction of the measured total injected
  energy, as function of the full jet length.}
  \label{fig:etherm-fraction}
\end{figure}
Figure~\ref{fig:etherm-fraction} shows the increase in thermal energy as fraction
of the total injected power. Already for the heaviest jet (M1), most
of the injected (kinetic) power appears as thermal energy due to compression and
irreversible entropy generation at shocks. The thermal fraction increases not
only with time, but is also much stronger for the lighter jets, where
a thermalization of $\ga 80$ per cent is reached. Half of the thermal energy gain 
is found in the cocoon and half in the (shocked) ambient gas.
\citet{ONeill+2005} find $\approx 40$ per cent of the jet power in the thermal
ambient gas for their 3D jets with density contrast $0.01$ in a uniform
atmosphere, while in our simulations we find $\approx 35$ per cent, which is
quite good agreement.

\begin{figure}
  \centering
  \includegraphics[width=0.95\linewidth]{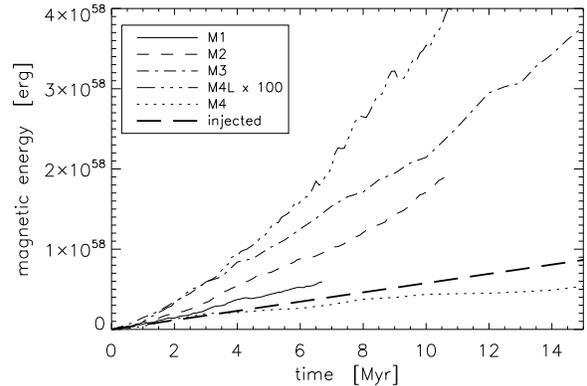}
  \caption{Evolution of the magnetic energy with time. The long-dashed line
  shows the injected amount of magnetic energy. For the case of M4L, magnetic
  energy was multiplied by 100 to account for the 10 times smaller field
  strengths.}
  \label{fig:magenergy-t}
\end{figure}
Magnetic energy only has a very small contribution (below 1 per cent), except for M4,
which is magnetically dominated and which has a magnetic energy contribution
rising up to 5 per cent. More than 90 per cent of the magnetic energy is located in the
cocoon. For all runs except M4, the magnetic energy that is actually measured is
significantly larger than the injected magnetic energy (this effect is stronger
for the lighter jets) and it grows faster then just linearly in time --
approximately with a power law exponent of $1.2$ {(Fig.~\ref{fig:magenergy-t})}.
Hence, other forms of energy seem to be converted into magnetic energy. For M4,
the measured magnetic energy is lower than expected from the nozzle values,
which may indicate that the additional thrust in the nose cone actually consumes
magnetic energy. 

The remaining fraction is kinetic energy, which is decreasing more and more with
lower jet density. Although the jet beam is the only energy input to the
system (at the nozzle), its contribution to the total kinetic energy is 10 per cent or
less, and 50 to 30 per cent is in the cocoon. The remainder, 50 to 70 per cent, comes from
the outward moving shocked ambient gas.

\subsection{Magnetic fields}

Magnetic fields are not only passive properties of the jet plasma, but an active
ingredient for the dynamics. One parameter describing this is the ratio between
thermal and magnetic pressure of the plasma ($\beta = 8 \pi p/B^2$). 
For the simulations described here, we used a fixed value for jet speed, Mach
number and magnetic field. Thus, the plasma $\beta$ cannot be
constant throughout the different runs (see Tab.~\ref{tab:simpara-var}). 
While M1 and M2 have passive magnetic fields, M3 and M4L have fields with
significant impact, and for M4 they are even dominant. 

The helical field configuration in the jet initiates an intriguing interplay
between kinetic and magnetic energy. Although the jet matter is injected without
any rotation, the Lorentz force from the helical field generates a toroidal
velocity component, as also found by \citet{Koessl+1990a}. This effect is
stronger for the runs with stronger magnetic fields (lower plasma $\beta$).  The
rotation does not originate from persisting angular momentum from the jet
formation, which should be very small due to the expansion of the jet. Also, it
is not continuous throughout the beam and even changing sign at some internal
shocks and interaction with cocoon vortices.

When the plasma reaches the terminal shock, it flows away from the axis radially
and turns back, forming the backflow that inflates the cocoon. Rough conservation of
angular momentum $l$ then produces a radially declining angular velocity $\Omega = l/R^2$
(differential rotation). 
Writing the induction equation in cylindrical coordinates, 
\begin{equation}
  \frac{\partial B_\phi}{\partial t} = -R \, \left(\vec{u}_\mathrm{p} \cdot \nabla\right)
  \frac{B_\phi}{R} - B_\phi \nabla \cdot \vec{u}_\mathrm{p} + R \,
  \left(\vec{B}_\mathrm{p} \cdot \nabla\right) \Omega, 
  \label{eqn:inductiontoroidal}
\end{equation}
it becomes evident that this shearing transforms poloidal field $B_\mathrm{p}$
into toroidal field $B_\phi$, also transferring kinetic energy into magnetic
energy, which explains why the contribution of magnetic fields to the total
energy is higher than its injected contribution.

We note that this creation of toroidal field in the jet head is not an artefact
of axisymmetry, but merely a consequence of allowing threedimensional vectors in
the simulation ($\vec{u}$ and $\vec{B}$). We do not expect this to be much
different in full 3D, apart from a naturally more complex structure in the
details.
What, in contrast, most probably is an artefact of axisymmetry is the
persistence of the toroidal field component in the cocoon. The cocoon plasma is
highly turbulent (Sect.~\ref{sec:cocoon}) with relatively little systematic motion, 
which is an intrinsically threedimensional phenomenon. This can
easily convert toroidal and poloidal field into one another, establishing some
dynamical equilibrium between those components, but maintaining the overall
field strength.

\begin{figure*}
  \centering
  \includegraphics[width=0.95\linewidth]{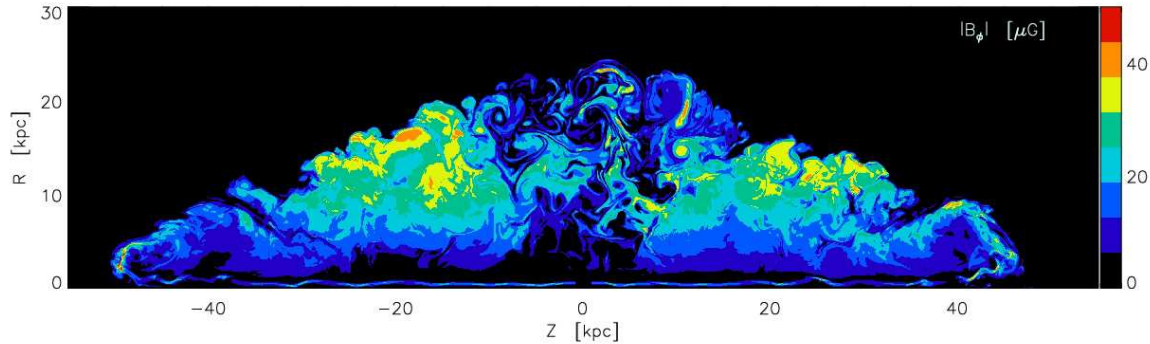}
  \caption{Toroidal magnetic field magnitude for M3 at 15 Myr. The right jet has
  positive toroidal field, the left jet has negative sign.}
\end{figure*}
\begin{figure}
  \centering
  \includegraphics[width=0.95\linewidth]{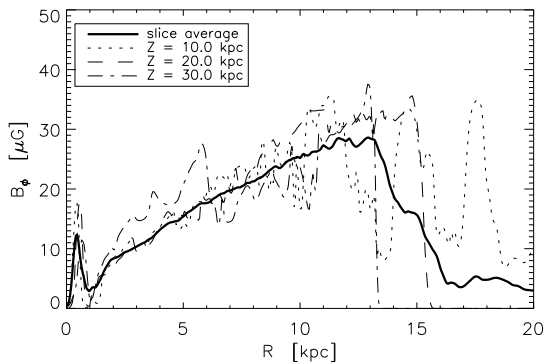}
  \caption{Toroidal field $B_\phi$ in three slices parallel to the midplane, and
  the average of the whole region $Z=10 \ldots 30$ kpc. Outside the jet radius
  $r_\mathrm{j} = 1$ kpc, the toroidal field shows a roughly linear increase
  with $R$.  Although in every slice $B_\phi$ drops to 0 at $R \approx 1$ kpc,
  the average does not drop to 0 (and thus is a bit misleading), as the
  positions are slightly offset for different slices.}
  \label{fig:bphi-slice}
\end{figure}
\begin{figure*}
  \centering
  \includegraphics[width=0.95\linewidth]{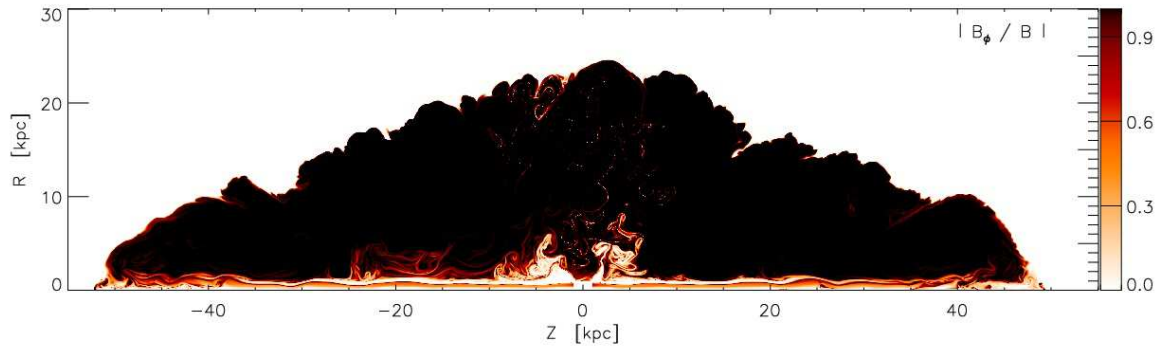}
  \caption{Fractional contribution of toroidal field to the total magnetic
  field. Simulation M3 at 15 Myr.}
\end{figure*}
\begin{figure}
  \centering
  \includegraphics[width=0.95\linewidth]{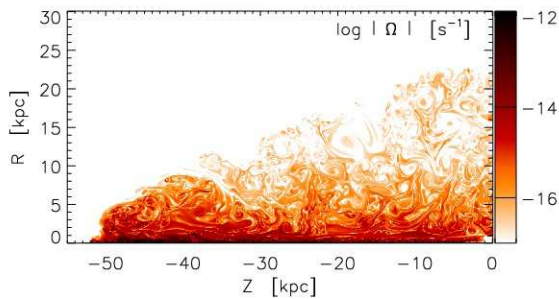}
  \caption{Angular velocity $|\Omega| = |v_\phi| / R$ in units of s$^{-1}$,
  scaled logarithmically. Note that there is a strong decline from the beam to the
  backflow.}
  \label{fig:omegamap}
\end{figure}
\begin{figure*}
  \centering
  \includegraphics[width=0.95\linewidth]{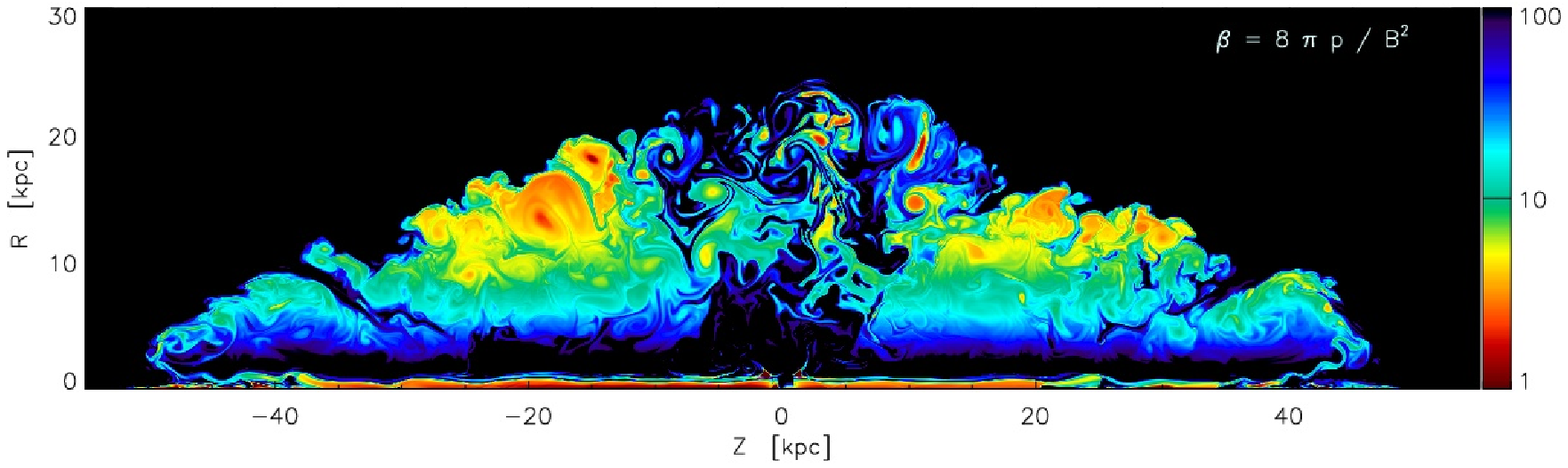}
  \caption{Plasma $\beta$ distribution for M3 at 15 Myr, in logarithmic scaling.}
  \label{fig:plasmabeta}
\end{figure*}

Comparing purely hydrodynamical models with the MHD models
(Fig.~\ref{fig:hydroMHDcomp}), we
find that the global properties, such as bow shock and cocoon sizes, are
generally robust if the magnetic fields are not dominant (as with M4).
The details, though, are different. While the hydro models show a ragged
contact surface between jet plasma and ambient gas due to Kelvin--Helmholtz (KH)
instabilities excited by the backflow, the MHD runs show a pronounced jet head,
which is clearly more stable, since the KH instability is damped by the magnetic
fields \citep[e.g.][]{MiuraPritchett1982}. Magnetic tension acts as a restoring force on the
growing instabilities, suppressing entrainment of ambient matter and ``fingers''
of dense gas reaching into the backflow, which is evident from the clearly lower
average density in the jet head region. The stabilizing effect appears at $\beta
\sim 10$ in the jet head. For the simulations with weaker fields there is no
noticeable difference between the magnetized and the pure hydrodynamics case.

Damping of the KH instability by magnetic fields, however, only works with the
field component parallel to the instability wave vector, which in turn means
that in axisymmetry only the poloidal magnetic field can damp the instabilities at
the contact surface. 

Although the earlier-mentioned shearing mechanism amplifies magnetic fields and
should therefore provide even more damping of KH instabilities, we cannot see
this effect further away from the jet head, because in axisymmetry the backward
reaction (toroidal to poloidal) cannot work and thus the poloidal component
becomes too weak (Fig.~\ref{fig:lic-maps}).
As the magnitude of the magnetic field in the cocoon is as strong as in the jet
head, it seems reasonable that with balanced magnetic field components in
reality, the contact surface could be stabilized.

\begin{figure*}
  \centering
  \includegraphics[width=0.95\linewidth]{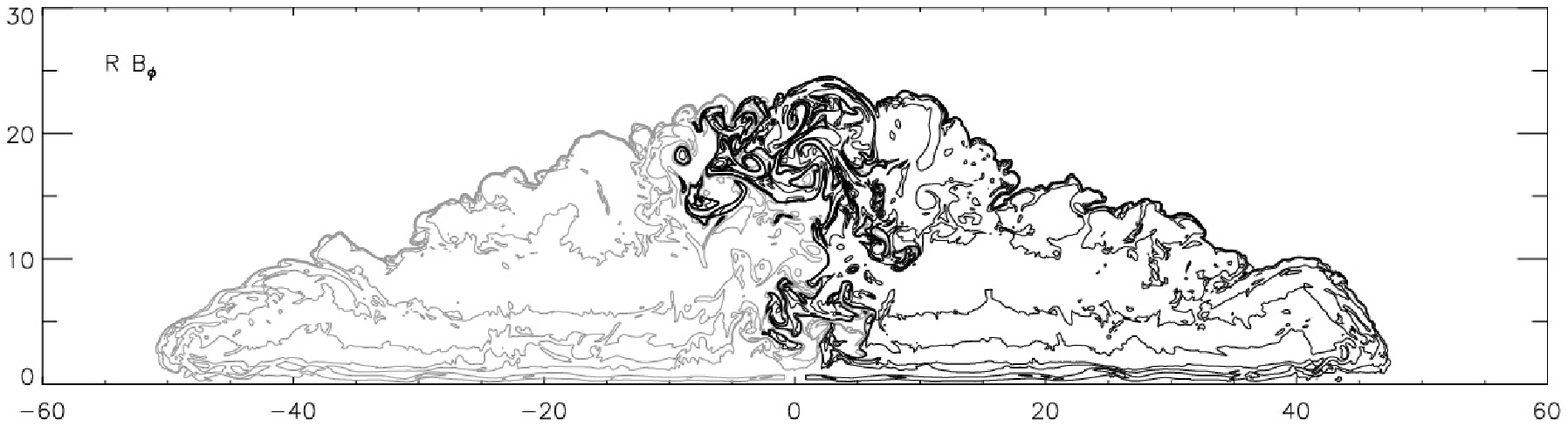}
  \caption{Poloidal current field lines (contours of $R B_\phi$). Grey:
  negative values, black: positive values. Model M3 at $t=15$ Myr.}
  \label{fig:RBphi}
\end{figure*}
The toroidal field $B_\phi$ is directly related to the generating current
$\vec{j}_\mathrm{p}$, which is shown in Fig.~\ref{fig:RBphi} as field lines. 
Our toroidal field setup describes a situation where the poloidal currents leave
the nozzle axially in the jet core, turning back in the sheath. As the backflow
develops, the poloidal current flows along the contact surface with typical
integrated currents of several $10^{18}$ amperes
\citep{Camenzind1990,Blandford2008}. The toroidal field in the cocoon, built-up
by the shearing in the jet head, seems to form its own current circuits.  The
gross radial behaviour $B_\phi \propto R$ (Fig.~\ref{fig:bphi-slice}) can be
attributed to the relatively uniform distribution of the axial current through
the planes perpendicular to the jet beam. 

If the toroidal field is strong in the jet head region, the Lorentz force
$f_\mathrm{L} = \vec{j} \times \vec{B} / c$ produces additional thrust for the
jet propagation due to the strong radial current component, which is evident for
M4, showing a pronounced nose cone, and may also explain the slightly faster
propagation of M3 with respect to H3 (Fig.~\ref{fig:colenwidth}).

\begin{figure*}
  \centering
  \includegraphics[width=0.95\linewidth]{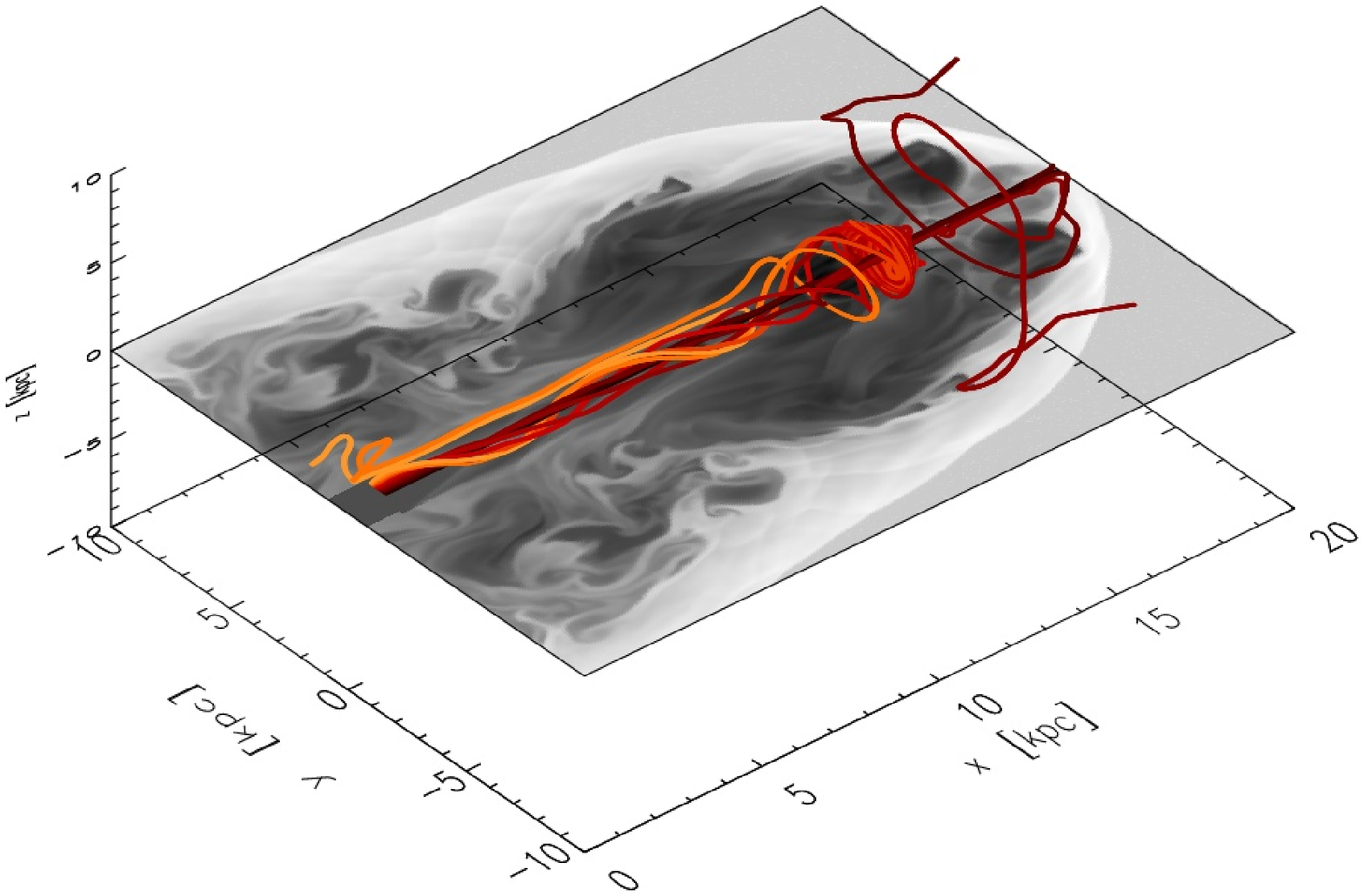}
  \caption{3D magnetic field lines from within the beam with a transparent
  grey-scale logarithmic density slice in a plane through the jet axis. Model M2
  at $t = 1.3$ Myr.}
  \label{fig:fieldtrace}
\end{figure*}
Inside the beam, the magnetic field stays mostly poloidal, as injected, but near
the terminal shock it is compressed axially, directed off the axis and sheared,
producing strong toroidal field loops (Fig.~\ref{fig:fieldtrace}).

\begin{figure}
  \centering
  \includegraphics[width=0.95\linewidth]{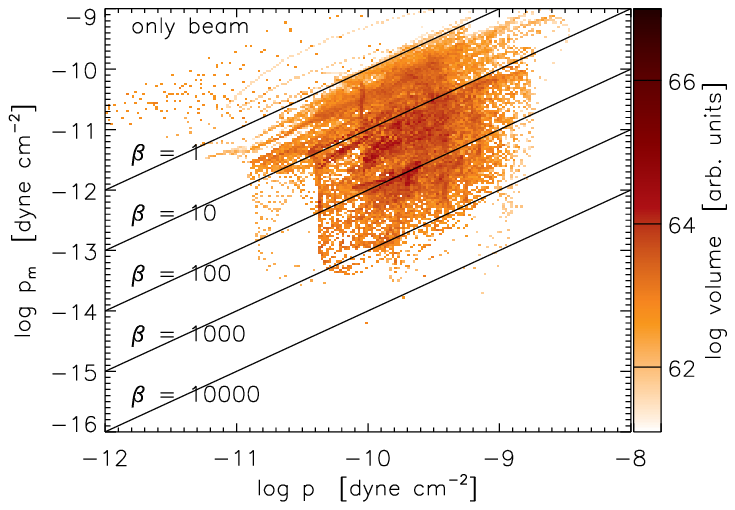}
  \includegraphics[width=0.95\linewidth]{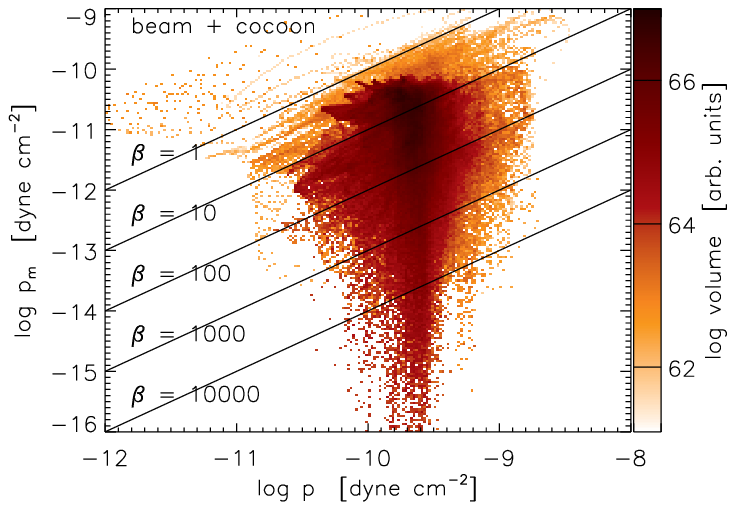}
  \caption{$p_\mathrm{m}$--$p$ histogram for the beam (defined by kinetic energy
  flux $\ge 1$ \% of the maximum value) and all the jet matter from simulation
  M3 after $15$ Myr. Lines of constant plasma $\beta$ with values of 1, 10, 100,
  1000 and 10\,000 are overlaid.}
  \label{fig:pmag-p}
\end{figure}

Finally, we turn to volume-weighted 2D histograms of magnetic
pressure $p_\mathrm{m} = B^2/8\pi$ and thermal gas pressure $p$ in
Fig.~\ref{fig:pmag-p}, where the contributions from only the jet beam
and all the jet plasma is shown separately.

The jet nozzle is located at $(\log p,\log p_\mathrm{m}) \approx (-10,-10.5)$ as a
vertical line (constant pressure, but radially varying magnetic field). As the
matter flows through the beam, internal shocks (cf.
Figs.~\ref{fig:etacomp-pressure} and \ref{fig:hydroMHDcomp}) cause strong
changes in pressure whereas the plasma $\beta$ remains unchanged (magnetic field
is compressed with the plasma), leading to lines originating from the nozzle
location parallel to the overplotted $\beta = \mathrm{const}$ lines. The plasma
$\beta$ somewhat increases along the beam when it it interacts with the cocoon
vortices, and thus creates some down-shifted parallels. 
There is no clear separation between the beam and the enclosing cocoon in the
beam-only diagram, hence both shear layers of the beam and cocoon gas are contained in the
wide area below $\beta = 10$. Still, there are strong pressure changes indicated
by the wide horizontal distribution.

The distribution of cocoon cells widely spreads both to higher and lower magnetic
fields from this area. The pronounced trail downwards is the transition to the
ambient gas through entrainment; since the ambient gas is essentially
unmagnetized, it is located even below the lower border of the figure
(Fig.~\ref{fig:pmag-p}). The radial increase of magnetic field in the cocoon
yields the extension towards lower plasma $\beta$ (see also
Fig.~\ref{fig:plasmabeta}). The spiky features around $\beta ~\sim 2$ are single
vortices in the outer parts of the cocoon, where the pressure drops towards the
centre due to centrifugal forces together with a slight increase of toroidal
field. Altogether, the spread of the cocoon cells is considerably larger in magnetic
pressure than in thermal pressure.
\begin{figure}
  \centering
  \includegraphics[width=0.95\linewidth]{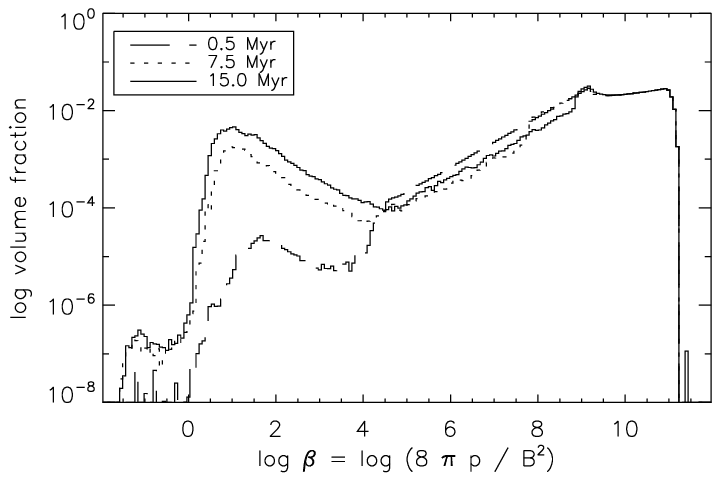}
  \caption{Volume-weighted histogram of plasma $\beta$ for M3 at $0.5$, $7.5$
  and $15$ Myr.}
  \label{fig:betahist}
\end{figure}

The situation shown in Figs.~\ref{fig:pmag-p} and \ref{fig:betahist} is typical
for the time evolution of these diagrams. Clearly, some features are appearing,
changing and disappearing continuously, such as individual internal shock lines
or the cocoon vortex spikes. The general structures in the diagrams persist at
all times. There are, however, two systematic changes with time: Firstly, the
``cocoon bump'' in Fig.~\ref{fig:betahist} ($\log \beta \sim 1$) grows due to
cocoon expansion, eroding the ``ambient bump'' ($\log \beta \sim 9$), and moves
to the left, faster at early times and then becoming continuously slower.
Secondly, as the cocoon pressure drops, the cocoon distribution of
Fig.~\ref{fig:pmag-p} moves towards the left (and somewhat down due to the
mostly constant distribution in $\beta$ at late times), and grows
with cocoon volume, too.

\subsection{The lightest jets}
\label{sec:lightestjet}

The lightest jet in the series, M4, shows a very different behaviour from the
other runs due to its strong magnetic fields, thus a run with lower magnetic
fields (M4L) was performed in addition. In this subsection, we focus on the
specific properties of and differences between these two runs.

Both simulations show unstable beams, which are temporally stopped,
deflected or disrupted. This is quite natural for the very light jets, where the
impact of cocoon vortices hitting the beam is stronger, when the beam shows
lower inertia but the cocoon gas is dense due to entrainment and mixing with
the dense ambient gas. This destabilization is particularly strong in
axisymmetry, as the vortices cannot ``miss'' the beam as they could in 3D.
For M4L, after a strong deflection of the right beam near the nozzle ($t
\approx 20$ Myr), a small region with strong poloidal field piles up just next
to the nozzle and creates a magnetic layer ($\beta \ga 1$) at the beam
boundary. This protects the beam from cocoon vortices and entrainment, and from
there on inhibits disruptions of the right jet, which then propagates more
quickly than the left jet. At the end of the simulation, the right jet
is $\approx 20$ per cent longer than the left jet and shows an almost undisturbed
beam up to the jet head. More detailed examination of this phenomenon may be
interesting, but as it was only introduced by chance, the details are difficult
to reproduce and beyond the scope of this paper. None the less, the overall
propagation of the jet within the simulated time (Sect.~\ref{sec:cocoon}) is
not much affected by this.

Keeping the jet speed and the Mach number fixed, the ratio of the thermal
pressures of ambient gas and jet nozzle changes with density contrast, yielding
an underpressured jet for M4 and M4L.
For M4, the magnetic field in the nozzle is already stronger than equipartition
and the Alfv\'en speed is higher than the sound speed. This run is dynamically
dominated by the magnetic field and shows a pronounced nose cone, which is known
for jets with strong toroidal fields \citep{Clarke+1986}. Magnetic tension
pinches the jet matter into an narrow tube of $2.5$ to $3.5$ kpc radius,
completely suppressing a backflow and thus preventing the formation of a wide
cocoon. The simple case of a plasma column in radial magnetostatic equilibrium
keeps $p + {B_\phi^2}/{4 \pi}$ constant. If ${B_\phi^2}/{4 \pi}$ approaches the
thermal pressure, the magnetic pinch becomes important. In our case, the
toroidal field in the plasma column is relatively homogeneous, showing a
(volume-weighted) distribution mostly between $30$ and $50 \, \umu\mathrm{G}$,
while the thermal pressure lies (radially decreasing) in the range $1 \ldots 3
\times 10^{-10} \, \mathrm{dyne/cm^{2}}$, thus matching $p \sim B_\phi^2 / 4
\pi$ and being just around equipartition.  These values are not the ones set by
the jet nozzle, although those obey $p \sim B_\phi^2/ 4 \pi$, too. The twisting
and shearing processes described in the previous subsection are very strong due
to the equipartition-level magnetic fields, the rotation around the jet axis can
make up a large fraction of the total velocity, and the toroidal field component
grows to the measured values.

\citet{KrauseCamenzind2001} examined the convergence of a nose cone simulation and found
that the Mach disc retreated towards the nozzle and thus did not converge. 
Also in M4, the Mach disc is very near to the jet nozzle, and the velocities
after that shock are subsonic (although the nose cone itself propagates faster
than the jet head in M4L). Thus, it is unclear, how reliable the run is. We also note
that the magnetic pinch is subject to MHD instabilities \citep{Clarke1993}, which
might produce blobs and disrupt the plasma column in 3D. However, as this nose
cone is produced by the magnetic tension of the strong toroidal field, this is
not applicable to strong poloidal fields, which cannot provide the necessary
hoop stress, although it seems difficult to maintain a strong poloidal field
along an interacting beam without converting part of it into toroidal field,
which then might again pinch the plasma.

\section{Discussion}

\subsection{Magnetic Fields}

Effects of magnetic fields naturally depend on their strength. Trying to
understand the smoothness of jet cocoons in galaxy clusters, we concentrated on
magnetic fields in jets which are not dominant, but still have significant effects on
the jet dynamics, the best example for this being the M3 run with average plasma
$\beta = 8$. It is well known \citep{MiuraPritchett1982} that magnetic tension can damp or
suppress Kelvin--Helmholtz (KH) instabilities and hence it may be the key to
stabilizing the contact discontinuity between jet and ambient gas. However, how
this applies to the complex case of jet--ambient interaction is not yet known. 

We emphasize that much care was taken to use a globally consistent setup for our
very light jets, in particular: keeping the bow shock inside the computational
domain at all times; simulating bipolar (back-to-back) jets to remove an
artificial boundary condition in the midplane and allow interaction of the
backflows for a realistic lateral expansion; and using a configuration which
confines the magnetic field to the jet and has closed field lines instead of a
homogeneous magnetic field reaching to infinity, which is then effectively
anchored in the ambient gas. The assumed simplifications, axisymmetry and a
constant ambient density, make extraction of the underlying physics easier and
effects of relaxing those for hydrodynamic jets were previously investigated by
\citet{KrauseVLJ2}. Thus we expect to at least qualitatively model the
situation realistically. 

Two main effects arise from the inclusion of magnetic fields: Firstly, in the
jet head, we see that the provided magnetic fields in the jet do indeed
stabilize the contact surface, which produces a pronounced jet head and lobes,
similar to the ones seen in Cygnus A \citep{Carilli+1991} and other classical
double radio sources. Effects from an ambient density profile can be excluded 
due to the prescribed constant density atmosphere. Furthermore, the entrainment
of ambient gas is significantly smaller there than without magnetic fields.

Secondly, jets prove to be efficient generators of magnetic energy, transferring
part of their huge kinetic power to magnetic fields through shearing in the jet
head. This relies on some rotation of the beam plasma, which will (as seen in
the simulations) generally be present for a non-zero toroidal field component.
Some toroidal field is expected if the mostly axial field in the beam
\citep{BridlePerley1984} is perturbed three-dimensionally and from jet formation
models, where the toroidal field is necessary for jet collimation at least at
small scales. The shearing mechanism provides a source of magnetic energy for
the cocoon and furthermore affects the magnetic field structure at the hotspots
and possibly some internal shocks.  A radial and toroidal field component in the
beam is known to be compressed by the terminal shock and is then visible as a
strong magnetic field perpendicular to the jet axis. The jet head shearing
provides another mechanism, independent from compression, to greatly enhance the
toroidal field and thus produce a perpendicular field component stronger than
that expected from compression. For jets pointing more towards the observer, the
toroidal field around the hotspot region may become observable. 

This may be relevant for several observational findings, one being the
smoothness of radio cocoons. We have shown that even if the plasma $\beta$ is of
order ten, only, the fields in the backflow and the cocoon respectively will be
strong enough to damp KH instabilities at the cocoon--ambient gas interface and
yield a morphology much smoother than seen in hydrodynamic simulations,
reconciling simulations with observations of sources as Cygnus A
\citep{Lazio+2006}, Pictor A \citep{Perley+1997} or Hercules A
\citep{GizaniLeahy2003}, where the latter seems to be a past high-power source.
Due to the 2.5D nature of the simulations, the effect is restricted to the jet
head region. In a full 3D simulation, we expect therefore the cocoon--ambient
interface to be more stable even further back from the hotspots.
The amplification of beam magnetic fields in the ``jet head
machine'' furthermore is consistent with the observation of magnetic fields in
the cocoon just somewhat below equipartition
\citep{HardcastleCroston2005,Migliori+2007}. Additionally, the magnetic field
predominantly perpendicular to the jet axis in weak FR I sources might be
related to the expansion of the jet, which by the shearing would create strong
toroidal fields in the absence of strong turbulence. Even though the beam
rotation can change much due to interaction with the cocoon and shocks and even
change sign, the helicity of the toroidal field is not changed and can
thus link the field at large scales with the field topology near the black hole
\citep{Gabuzda+2008}.

For the magnetic field topology in the rest of the cocoon, axisymmetry is a major
limitation, contrary to the effects discussed before. Magnetic field in a
toroidal configuration cannot damp KH instabilities in axisymmetry since no
magnetic tension is available as restoring force, while poloidal field could do
so. Fortunately, the jet head-generated toroidal field in the turbulent cocoon
partly would be converted into poloidal field in three dimensions, establishing
some dynamical equilibrium between the components but keeping the overall field
magnitude or amplify it even further, and this makes the cocoon magnetic field a
reasonable explanation for the smooth contact surfaces. As a future step we will
examine this effect in three dimensions to be able to quantify the amount of
damping and suppressed entrainment of ambient gas in the cocoon.

However, despite the inability to actually produce the expected smooth contact
surfaces in axisymmetry away from the jet head region, there is no reason to
assume that the amplification of magnetic fields should be in three dimensions
any different than shown in our simulations, as the plasma dynamics is not
very different and the shearing mechanism in the jet head simply relies on the
off-axis flow of plasma, which also happens in 3D. Furthermore, we are not aware
of any reason that the field magnitude in the cocoon should be much different in
3D. It is unclear, though, how the spatial distribution of the magnetic field
would look like: 3D turbulence might want to distribute field strength rather
uniformly in the cocoon, but formation of a large-scale poloidal current may try
to establish a radially increasing toroidal field. Observations indicate that
magnetic field strengths within the cocoon may vary considerably
\citep{Goodger+2008}.

It may be interesting to note that the amplification of magnetic field is quite
related to dynamo action as in the sun. The shearing ($\Omega$ effect) is just
the same and solar convection is replaced by jet-driven cocoon turbulence, but
the location of these actions are different and they are externally powered (by the
beam thrust) instead of self-sustained. The spatial separation of the two
effects and the (at least roughly) isotropic turbulence, however, prevent the
formation of an outstanding large-scale poloidal field.

The uncertainty in the magnetic field topology in the cocoon also applies to the
distribution of plasma $\beta$ in the system. We (expectedly) found that plasma
$\beta$ is unchanged throughout shocks despite a gradual increase along the beam
(which might also be due to limited resolution of the beam and entrainment). 
Thus the assumption of a fixed fraction of equipartition to generate 
synchrotron emission maps from hydro simulations seems to be quite justified. 
However, this was not found to be true for the cocoon, where a wide distribution 
of $\beta$ was found and deriving synchrotron emission from hydro models 
thus may be far away from MHD results. But as mentioned, this result is expected to change in 3D, apart
from having relatively low $\beta$ in the cocoon. Emission maps of our
simulations and comparison to hydro models are beyond the scope of this work
and will be presented in a subsequent paper.

The amplification of magnetic fields is also particularly interesting for the
question of the origin of lobe magnetic fields. \citet{DeYoung2002} pointed out
that equipartion fields in the lobes cannot be passively advected with the
plasma from the jet beam due to flux conservation arguments. The beam magnetic
fields would have to be of order $0.01$ G or higher, certainly above
equipartion, which would result in enormous synchrotron losses, luminosities
incompatible with observational limits and probable disruption of the jet due to
the magnetic pressure. Hence, the magnetic field must be amplified by some
mechanism, and \citeauthor{DeYoung2002} argues for turbulent amplification in
the hotspot flow, though it is not easy to meet the necessary requirements for
this. The shearing in the jet head, which is seen in our simulations, in
contrast, almost inevitably provides this amplification and can therefore explain the
strong lobe magnetic fields or at least contribute to their field strength. In
fact, the simulations exhibit field magnitudes in the cocoon that are
comparable to field magnitudes in the beam and consequently have similar plasma
$\beta$ since the beam and cocoon pressures came to balance. We conclude that
shearing due to off-axis flow of the plasma provides a natural explanation for
the lobe magnetic fields and allows equipartition jets to inflate an
equipartition cocoon.

\subsection{Dynamical Evolution}

X-ray observations of the ambient cluster gas contain valuable
information about several jet and AGN properties and self-similar models 
can give easy access to underlying physical parameters. In the present paper, we 
are able to confirm agreement of our numerical simulations with self-similar models
\citep{Falle1991,Begelman1996,KaiserAlexander1997,KomissarovFalle1998} for the bow shock
propagation. 
Excentricity of the bow shock and its Mach number seem to be an easy way to 
compare theoretical models with observations, without the need for uncertain
assumptions on the emission of the radio plasma.

The weak and roundish bow shocks in observations indicate that models of very
light jets (with density ratios $< 10^{-2}$) are necessary for most cluster
sources. Although we chose a simplified setup with a constant ambient gas and
axisymmetry, the simulations are in the regime of observed values for various
sources and self-similar models generalize this behaviour for declining
cluster profiles, which was already examined for very light hydrodynamic jets by
\citet{KrauseVLJ2}. As our runs, with the clear exception of the magnetically
dominated M4, propagate as their hydro counterparts, only minor deviations from 
those results are expected, except where specific source properties are to be 
included.

Contrary to the bow shocks and the jet length, we find that the cocoon width
in general does not evolve self-similarly but for lighter jets grows with lower power
law exponents and the mean cocoon pressure drops more slowly than expected.
Although this may seem unexpected, it was already stated by
\citet{KaiserAlexander1997}, that contrary to the bow shock, the self-similar
evolution of the cocoon depends on the physical model for the post-hotspot flow
and thus, deviations are to be expected if these assumption do not hold in
the simulations. Since very light jet cocoons are less overpressured and approach 
the ambient pressure sooner, the sideways expansion becomes slower and may even stall, 
letting their aspect ratio (length to width ratio) grow. {This is in fact
observed by \citet{Mullin+2008}, who find a wider range of aspect ratios, once
the source size approaches $\sim 100$ kpc.} Similar behaviour
would be expected for the heavy jets, although at much later times. Thus, cocoon
evolution depends sensitively on the question of overpressure, which can be
addressed by the strength of the lateral bow shock. 
Self-similar models, in contrast, assume that the ambient pressure is
negligible. \citet{KomissarovFalle1998} defined two scales, $l_c$ and $L_c$,
between which they found self-similar evolution. The lower bound $l_c$, where
the swept-up mass equals the jet mass, is much smaller than the jet radius in
our simulations, and the upper bound $L_c$, where the ambient pressure becomes
important, is comparable to our computational domain size.  Accordingly, while
they observe the self-similarity being established, we observe its end,
explaining why our less overpressured numerical solutions gradually deviate from
a self-similar evolution. 

Furthermore, cluster density
profiles make cylindrical cocoons rather than elliptical ones due to the
weaker density contrast at larger distances \citep{KrauseVLJ2}. Altogether, this
makes us confident that our simulations reasonably well describe observed
cluster sources.

In contrast to bow shocks, measurements of the cocoon shape are
complicated by cooling of the relativistic electrons, which limits observations
to the outermost parts (lobes). While radio observations show the high-energy
electrons in the cocoon as lobes, single-fluid MHD simulations only trace the
low-frequency emitting matter and can only show the low-frequency radio morphologies
\citep[cf. high and low frequency images in][]{Carilli+1991}, which generally
suffer from low spatial resolution. This situation fortunately will much improve
in the future with new telescopes as LOFAR or the SKA, which will allow more
detailed studies of cocoon dynamics and turbulence. Until then, X-ray images of
cavities and (in some cases) the inverse-Compton emission off the cosmic
microwave background may supplement available low-frequency radio maps.

\citet{Scheuer1982} introduced the ``dentist's drill'' to refer to a moving working
surface, which therefore widens the jet head and the lobes. Very light jets naturally 
show extensive cocoons and varying deflection of the beam widens the jet head and 
hence, even in axisymmetry, show something very similar to a ``dentist's
drill''. While this does not exclude beam precession
\citep{SteenbruggeBlundell2008}, it is does not require it and no large
precession amplitudes are needed a priori. 

We expect for multiple outbursts of different power in the same cluster,
indicated by ``ghost cavities'' \citep[e.g.][]{Fabian+2006,Wise+2007}, that their
evolution crucially depends on the history of the past outbursts, as these push
the dense cluster gas aside, letting the new outburst propagate with different
density contrast. In this case, the new jet might quickly push forward to the
old jet size, then resuming its work on the dense ambient gas. The morphology of
the cavities may allow the determination of the respective density contrasts and
thus could shed light on the outburst history. 

The thermal interaction of jets with the intra-cluster medium is less accessible
to direct comparison with observations. Slower jet {head} propagation is
responsible for the strong impact of the beam at the working surface and a high
thermalization; some conversion of kinetic to thermal energy will additionally
occur near or in the beam due to beam destabilization, but may be less effective
in 3D. Despite the dominant power source is the kinetic jet power, this strong
thermalization converts most of the input power to thermal energy -- about half
of this in the shocked ambient gas and half in a cocoon filled with high-entropy
plasma, which eventually may transfer at least part of its energy to the
entrained cluster gas. This is in line with findings of other authors
\citep[e.g.][]{Reynolds+2002,ONeill+2005,Zanni+2005}, where the latter authors
conclude that up to 75 per cent of the energy can be dissipated irreversibly and
thus is available for heating in the intra-cluster medium, as required by the
X-ray luminosity--temperature relation \citep{MagliocchettiBrueggen2007}
 and to provide ``radio-mode'' feedback for models of galaxy evolution 
\citep{Croton+2006}.

Since only the hot gas phase is simulated, effects on the cold or warm phases of the 
interstellar medium (ISM) of galaxies are difficult to estimate. Clearly, the thermalization
efficiency cannot be simply applied to the cold gas. Simulations of multi-phase 
turbulence in the jet cocoon by \citet{KrauseAlexander2007} with their higher spatial 
resolution can resolve the different phases and provide a complementary view 
(``microphysics'') onto the jet--cloud interaction. However, even if the thermal 
energy mostly is deposited on the hot gas phase (at larger distances), 
it is evident from our simulations that the jet cocoon is a rich reservoir of 
turbulent kinetic energy which will act on the cold gas phase of the galaxy 
over a time scale corresponding to the decay time scale of the cocoon turbulence.
For a jet of power $10^{46}$ erg s$^{-1}$ being active for $10^7$ years, 
the turbulent energy stored in the cocoon is expected to be on the order of 
a few $\times 10^{59}$ ergs and over a time possibly longer than the jet activity
will interact with the cold ISM phases.

Another interesting result of the present simulations is the excitation of sound
waves in the ambient gas by vortices in the turbulent cocoon, which is more
effective for the very light jets with their extended cocoons. Vortex shedding
\citep{Norman+1982} quasi-periodically occurs in the jet head, and the vortices
then are advected with the backflow into the cocoon and provide an intermittent
source for the turbulent cascade, producing pressure waves. Waves like these are
visible in the Perseus cluster \citep{Fabian+2006,ShabalaAlexander2007} and,
although being hard to observe, may be an ubiquitous feature in galaxy clusters
with current or past jet activity. Their typical wave length might yield a link
to jet dynamics and cocoon turbulence. In the lightest of our jets (M4L), the
bow shock is just about to turn into a sound wave and then simply would join the
enclosed sound waves. Viscous damping may be a mechanism to reduce the
amplitudes in addition to the growing wave area and is another candidate for
preventing cooling flows \citep{Fabian+2005}, but in our scenario would be 
related to the jets rather than to the AGN itself.

Axisymmetry naturally imposes some constraints on the dynamics, which have to be 
considered carefully. Jet beams in high-power sources are essentially
axisymmetric objects and effects of the full third dimension are merely perturbations
from axisymmetry. However, this obviously is not true when beam 
stability or non-axisymmetric effects are explored specifically. While generally 
3D jets are subject to a greater number of instabilities, for very light jets there
is an opposing effect of an increased number of dimensions. While in 3D, cocoon vortices 
often will miss the beam or are slightly deflected, this is not possible in
axisymmetry and the beam thus is destabilized, deflected or disrupted more easily
which is most evident from our lightest run (M4L). As seen in the very light jets
of \citet{KrauseVLJ2}, the beam stability improves when going to full three 
dimensions. 
For most results, however, energetics and scaling behaviour are not expected to
change significantly in 3D, notable exceptions to this being cocoon turbulence,
magnetic field topology and stability of the contact discontinuity.

Cocoon turbulence further away from the jet head certainly will differ with
increased dimensionality as the increased number of degrees of freedom for
vortices allow them to turn in all directions and interactions between colliding 
vortices will be different. Though, we expect the effects on cocoon morphology
to be within reasonable limits, as the kinetic energy in the cocoon is lower
than the thermal energy by factors of $\ga 3$ for $\eta \le 10^{-2}$ and hence,
effects of thermal pressure will dominate.

\section{Conclusions and Summary}

We performed a series of axisymmetric hydrodynamic and magnetohydrodynamic 
simulations of bipolar very underdense jets in a constant density atmosphere. 
The magnetic field is mostly confined to the jet with a helical topology.

(1)
We find that the magnetic fields damp Kelvin--Helmholtz instabilities in the
jet head and stabilize it. They produce smoother and more pronounced outer
lobes already with a plasma $\beta \sim 10$. The entrainment of ambient gas
into the cocoon is considerably suppressed there. This morphology is more
consistent with observations of powerful double radio sources than are
hydrodynamic simulations, which show a ragged cocoon boundary.

(2)
Magnetic fields are efficiently amplified in the jet head by shearing as the
plasma streams off the jet axis. This originates from a rotation of the beam
which we find to be a general result of a toroidal field component being
present (yet not necessarily dominant) in the jet. The shearing converts part of
the huge kinetic energy into magnetic energy and provides the cocoon with a
magnetic field much stronger than expected from flux conservation, in some
regions even approaching equipartition. These findings are consisistent with
recent observations of near-equipartition magnetic fields in cocoons derived
from radio/inverse-Compton emission observations. Already in our axisymmetric
simulations the fields are in principle strong enough to stabilize the contact
surface between the cocoon and the ambient gas all-over the cocoon and not only
in the jet head. The necessary change in field topology would be a consequence
expected of fully threedimensional turbulence in the cocoon.

(3) 
The amplified magnetic field is mostly toroidal, resulting in a stronger
contribution of the field component perpendicular to the jet axis than expected
from pure compression of magnetic fields at the hotspots. It is also expected
at locations where a jet widens considerably (as in FR I sources). In the
backflow and the cocoon, however, turbulence will probably establish some
balance between the magnetic field components, which could not be established
in axisymmetry.

(4) 
The very light jets show round bow shocks with low Mach numbers. We find that
the bow shocks evolve self-similarly and hence give a simple link between observations
and some underlying physical parameters. The cocoon width, however, evolves
self-similarly only for jets in their highly overpressured phases, but grows
slower as the cocoons approach pressure balance with the ambient gas and the
bow shock Mach number drops. These sources thus are surrounded by thick layers
of shocked ambient gas. 

(5)
The jet cocoon shows highly turbulent motion. It is driven by vortices shed in
the jet head, which are advected with the backflow. Interaction of these
vortices with the ambient gas excites waves and ripples in the shocked ambient
gas, which are joined by the dissolving bow shock at later stages.

(6)
The strong thermalization that occurs for very light jets transfers most of the
jet power to the thermal energy of the cocoon and the shocked ambient gas,
making it available for heating of the cluster gas and radio-mode feedback. 
In addition to this, the turbulent motion in the cocoon is associated with a 
considerable amount of kinetic energy ($\sim 10$ per cent of the jet power) 
that may provide efficient feedback onto the cold phase of the galaxy's 
interstellar medium.

\section*{Acknowledgments}

We thank Paul Alexander and Martin Hardcastle for helpful discussions as
well as the anonymous referee for suggestions that further improved this paper. 
This work was supported by the Deutsche Forschungs\-gemeinschaft 
(Sonderforschungsbereich 439). The simulations partly have been carried out on
the NEC SX-6 of the HLRS Stuttgart (Germany).

\bibliographystyle{mn2e}
\bibliography{references}

\label{lastpage}

\end{document}